\documentclass[%
 reprint,
 amsmath,amssymb,
 aps,
]{revtex4-1}

\usepackage{graphicx}% Include figure files
\usepackage{dcolumn}% Align table columns on decimal point
\usepackage{bm}% bold math
\begin{document}

\preprint{APS/123-QED}

\title{Brownian  motion of magnetic domain walls and skyrmions, \\and their diffusion constants}

\author{Jacques Miltat}
\email{jacques.miltat@u-psud.fr}
\author{Stanislas Rohart}
\author{Andr\'{e} Thiaville}

\affiliation{%
Laboratoire de Physique des Solides, Universit\'{e} Paris-Sud, Universit\'{e} Paris-Saclay, CNRS, UMR 8502, F-91405 Orsay Cedex, France
}%

\date{\today}% It is always \today, today,
             %  but any date may be explicitly specified

\begin{abstract}
Extended numerical simulations enable to ascertain the diffusive behavior at finite temperatures of chiral walls and skyrmions in ultra-thin model Co layers exhibiting symmetric - Heisenberg - as well as antisymmetric - Dzyaloshinskii-Moriya - exchange interactions. The Brownian motion of walls and skyrmions is shown to obey markedly different diffusion laws as a function of the damping parameter. Topology related skyrmion diffusion suppression with vanishing damping parameter, albeit already documented, is shown to be restricted to ultra-small skyrmion sizes or, equivalently, to ultra-low damping coefficients, possibly hampering observation. 
%\begin{description}
%\item[PACS numbers]
%May be entered using the \verb+\pacs{#1}+ command.
%\end{description}
\end{abstract}

\pacs{Valid PACS appear here}% PACS, the Physics and Astronomy
                             % Classification Scheme.
%\keywords{Suggested keywords}%Use showkeys class option if keyword
                              %display desired
\maketitle

%\tableofcontents

\section{\label{Introduction}Introduction\protect\\ }

The prospect of ultra-small stable information bits in magnetic layers in presence of the Dzyaloshinskii-Moriya (DM) interaction \cite{Heinze:2011} combined to the expectation of their minute current propagation \cite{Jonietz:2010}, notably under spin-orbit torques \cite{Sampaio:2010}, builds up a  new paradigm in information technology. In stacks associating a metal with strong spin-orbit interactions e.g. Pt and a ferromagnetic metal such as Co, that may host isolated skyrmions, large domain wall velocities have also been forecast \cite{Thiaville:2012} and observed \cite{Jue:2016}. The DM interaction induces chiral magnetization textures, walls or skyrmions, that prove little prone to transformations of their internal structure, hence their extended stability and mobility. 

In order, however,  to achieve low propagation currents, steps will need to be taken towards a reduction of wall- or skyrmion-pinning. Recent experimental studies indicate that skyrmions fail to propagate for currents below a threshold roughly equal to $2\:10^{11}\mathrm{Am^{-2}}$ for $\text{[Pt/Co/Ta]}_{\mathrm n}$ and $\text{[Pt/CoFeB/MgO]}_{\mathrm n}$ multilayers \cite{Woo:2016}, or $2.5\:10^{11}\mathrm{Am^{-2}}$ for $\text{[Pt/(Ni/Co/Ni)/Au/(Ni/Co/Ni)/Pt]}$ symmetrical bilayers \cite{Hrabec:2017}. Only in one seldom instance did the threshold current fall down to about $2.5\:10^{10}\mathrm{Am^{-2}}$ for a $\text{[Ta/CoFeB/TaO]}$ stack, still probably, however, one order of magnitude higher than currents referred to in simulation work applying to perfect samples \cite{Jiang:2016}.

In a wall within a Co stripe $50~\mathrm{nm}$ wide, $3~\text{nm}$ thick, the number of spins remains large, typically $ 2^{16} $ for a $5~\text{nm}$ wide wall. A skyrmion within a Co monolayer (ML) over Pt or Ir, on the other hand, contains a mere 250 spins, say $ 2^{8} $.
Assuming that a sizeable reduction of pinning might somehow be achieved, then a tiny structure such as a skyrmion is anticipated to become sensitive, if not extremely sensitive, to thermal fluctuations. 

In this work, we show, on the basis of extended numerical simulations, that both chiral walls and skyrmions within ferromagnets obey a diffusion law in their Brownian motion at finite temperature \cite{Einstein:1905, Langevin:1908}. The diffusion law is shown to be valid over a broad range of damping parameter values. The thermal diffusion of domain walls seems to have attracted very little attention, except for walls in 1D, double potential, structurally unstable, lattices \cite{Wada:1978}, a source of direct inspiration for the title of this contribution. Chiral magnetic domain walls are found below to behave classically with a mobility inversely proportional to the damping parameter. As shown earlier \cite{Schutte:2014, Troncoso:2014}, such is not the case for skyrmions, a behavior shared by magnetic vortices \cite{Kamppeter:1999}. Vortices and skyrmions in ferromagnetic materials are both characterized by a definite topological signature. In contradistinction, skyrmions in antiferromagnetic compounds are characterized by opposite sign spin textures on each sublattice, with, as a result, a classical, wall-like, dependence of their diffusion constant \cite{Barker:2016}. Lastly, ferrimagnets do display reduced skyrmion Hall angles \cite{Woo:2018}, most likely conducive to modified diffusion properties.
\begin{figure}
\centering
\includegraphics[width=0.42\textwidth]{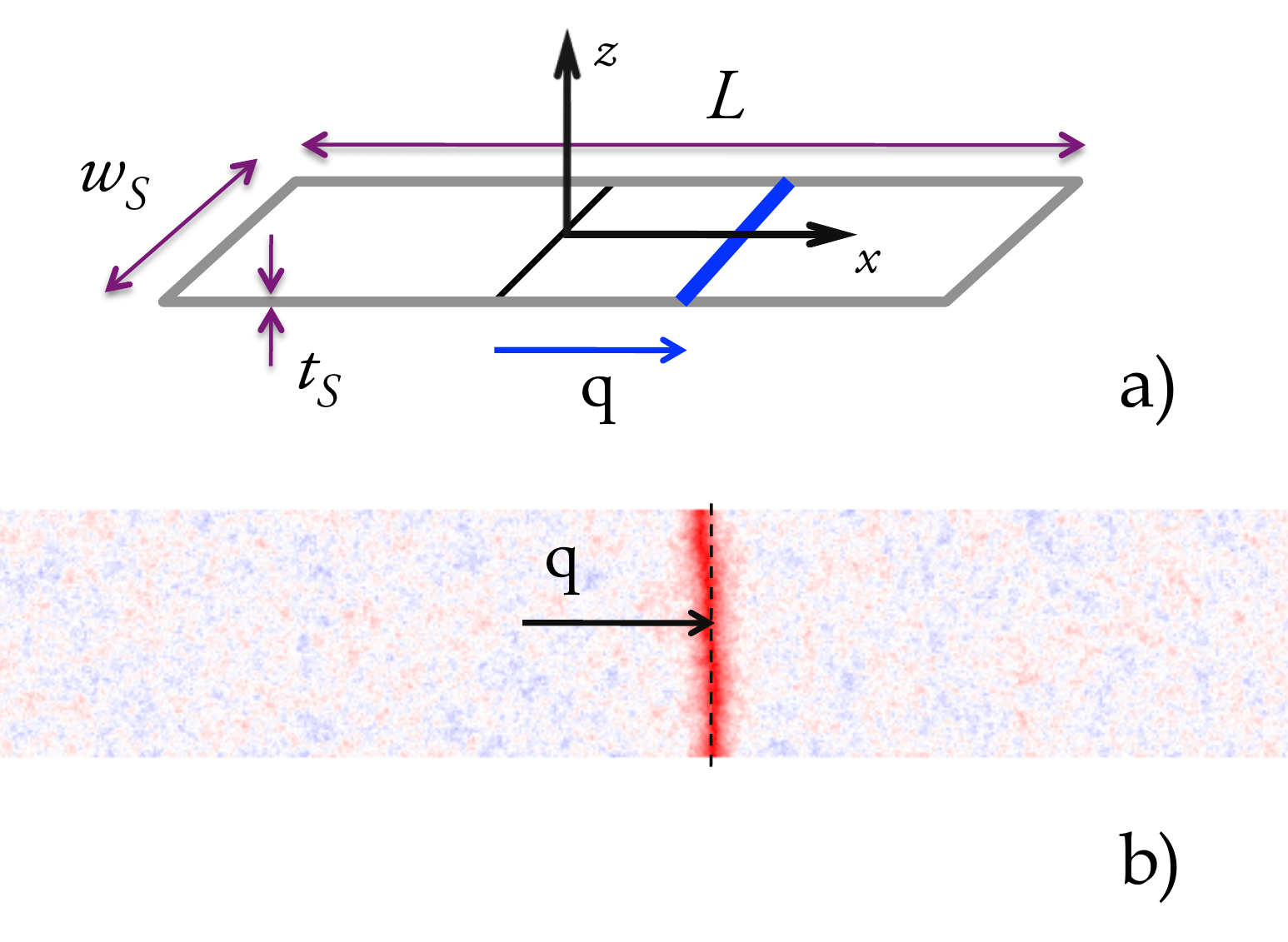}
\caption{a) Wall within a narrow stripe:  $w_\mathrm{S}$ is the stripe width, $t_\mathrm{S}$ its thickness. The stripe element length $L$ is solely defined for computational purposes.  $q$ is the wall displacement; b) snapshot of the magnetization distribution: color coding after $m_x$. The wall region $m_x \approx 1$ appears red. Thermal fluctuations are visible within domains: $T = 25~\mathrm{K}$,  $w_\mathrm{S} = 100~\textrm{nm}$, $t_\mathrm{S} = 0.6~\textrm{nm}$, $\alpha = 0.5$.}
\label{Fig_WSk_Diff_1}
\end{figure}
\section{\label{WallDiff}Domain wall diffusion\protect\\ }
We examine here, within the micromagnetic framework, the Langevin dynamics of an isolated domain wall within a ferromagnetic stripe with thickness $t_{\mathrm{S}}$, width $w_{\mathrm{S}}$ and finite length $L$ (see Fig.~\ref{Fig_WSk_Diff_1}). The wall is located at mid-position along the stripe at time $t = 0$. Thermal noise is introduced via a stochastic field $\vec H_{\mathrm {Rd}}$ uncorrelated in space, time and component-wise, with zero mean and variance $\eta$ proportional to the Gilbert damping parameter $\alpha$ and temperature $T$ \cite{Brown:1963} :
\begin{equation}\label{eq.WSkDiff:1}
\begin{split}
\langle \vec H_{\mathrm {Rd}} \rangle &=  \vec 0 \\
\langle  H_{\mathrm {Rd}}^{i} (\vec{r},t)  H_{\mathrm {Rd}} ^{j} (\vec{r'},t') \rangle &=
\eta \thinspace \delta_{ij} \thinspace  \delta(\vec{r}-\vec{r'}) \delta (t-t') \\
\eta &=\frac{2 k_{\mathrm B} T }{ \gamma_{0} \mu_0 M_{\mathrm S} } \alpha
\end{split}
\end{equation}
where, $k_{\mathrm B}$ is Boltzmann constant, ${\mu_0}$ and ${\gamma_0}$ are the vacuum permability and gyromagnetic ratio, respectively, $M_{\mathrm S}$ the saturation magnetization. Written as such, the functions $\delta(\vec{r}-\vec{r'})$ and $\delta(t-t')$ have the dimension of  reciprocal volume and time, respectively. Applied to numerical simulations, the variance of the stochastic field becomes $\eta =\frac{2 k_{\mathrm B} T }{ \gamma_{0} \mu_0 M_{\mathrm S} V dt } \alpha$, where $V$ is the computation cell volume and $dt$ the integration time step.
\subsection{Simulation results}
\label{sec:2a}
The full set of numerical simulations has been performed by means of an in-house code ported to graphical processing units (GPU's). Double precision has been used throughout and the GPU-specific version of the "Mersenne twister" \cite{Saito:2013} served as a source of long-sequence pseudo-random numbers generator. 

Material parameters have been chosen such as to mimic a 3-ML Co layer (thickness $t_\mathrm{S}=0.6~\text{nm}$) on top of Pt with an exchange constant equal to $A = 10^{-11}$~J/m, a $M_{\mathrm {s}} = 1.09\:10^{6}$A/m saturation magnetization, a $K_{\mathrm {u}} = 1.25\:10^{6}$~J/${\mathrm{m^3}}$ uniaxial anisotropy constant allowing for a perpendicular easy magnetization axis within domains, and a moderate-to-high DM interaction (DMI) constant $D_\mathrm{DM} = 2$~mJ/${\mathrm{m^2}}$. In order to temper the neglect of short wavelength excitations \cite{Berkov:2002}, the cell size has been kept down to $L_{\mathrm {x}} = L_{\mathrm {y}} = 1~\mathrm{nm}$, whilst  $L_{\mathrm {z}} = t_{\mathrm {S}} = 0.6~\mathrm{nm}$. The stripe length has been kept fixed at $L = 1~\mu \text{m}$, a value compatible with wall excursions within the explored temperature range. The  latter has, for reasons to be made clear later, been restricted to $\approx 1/3$ of the presumed Curie temperature for this model Co layer. Finally, the integration time constant, also the fluctuating field refresh time constant, has been set to $dt = 25~\text{fs}$.

As shown by the snapshot displayed in Fig.~\ref{Fig_WSk_Diff_1}b, the wall may acquire some (moderate) curvature and/or slanting during its Brownian motion. Because wall diffusion is treated here as a 1D problem, the wall position $q$ is defined as the average position owing to :
\begin{equation}
\label{eq:WSkDiff:2}
q = \frac{L}{N_{\mathrm {x}} N_{\mathrm {y}}} \frac 
{\sum_{i=1}^{N_{\mathrm {x}}}  \sum_{j=1}^{N_{\mathrm {y}}}  m_{\mathrm {z}} \left( i,j \right)  }  
{ \left[ \left\langle m_{\mathrm {z}} \right\rangle_{\mathrm {L}} - \left\langle m_{\mathrm {z}} \right\rangle_{\mathrm {R}} \right]}
\end{equation}
where, $i$ and $j$ are the computation cell indices, $N_{\mathrm {x}}$ and $N_{\mathrm {y}}$ the number of cells along the length and the width of the stripe, respectively, $\left\langle m_{\mathrm {z}} \right\rangle_{\mathrm {L}}$ is the fluctuations averaged value of the $z$ magnetization component far left of the domain wall, $\left\langle m_{\mathrm {z}} \right\rangle_{\mathrm {R}}$ the average value of $m_{\mathrm {z}}$ far right. Regardless of sign, $\left\langle m_{\mathrm {z}} \right\rangle_{\mathrm {R}}$ and $\left\langle m_{\mathrm {z}} \right\rangle_{\mathrm {L}}$ are expected to be equal in the absence of any $H_z$ field.
\begin{figure}
\centering
\includegraphics[width=0.40\textwidth]{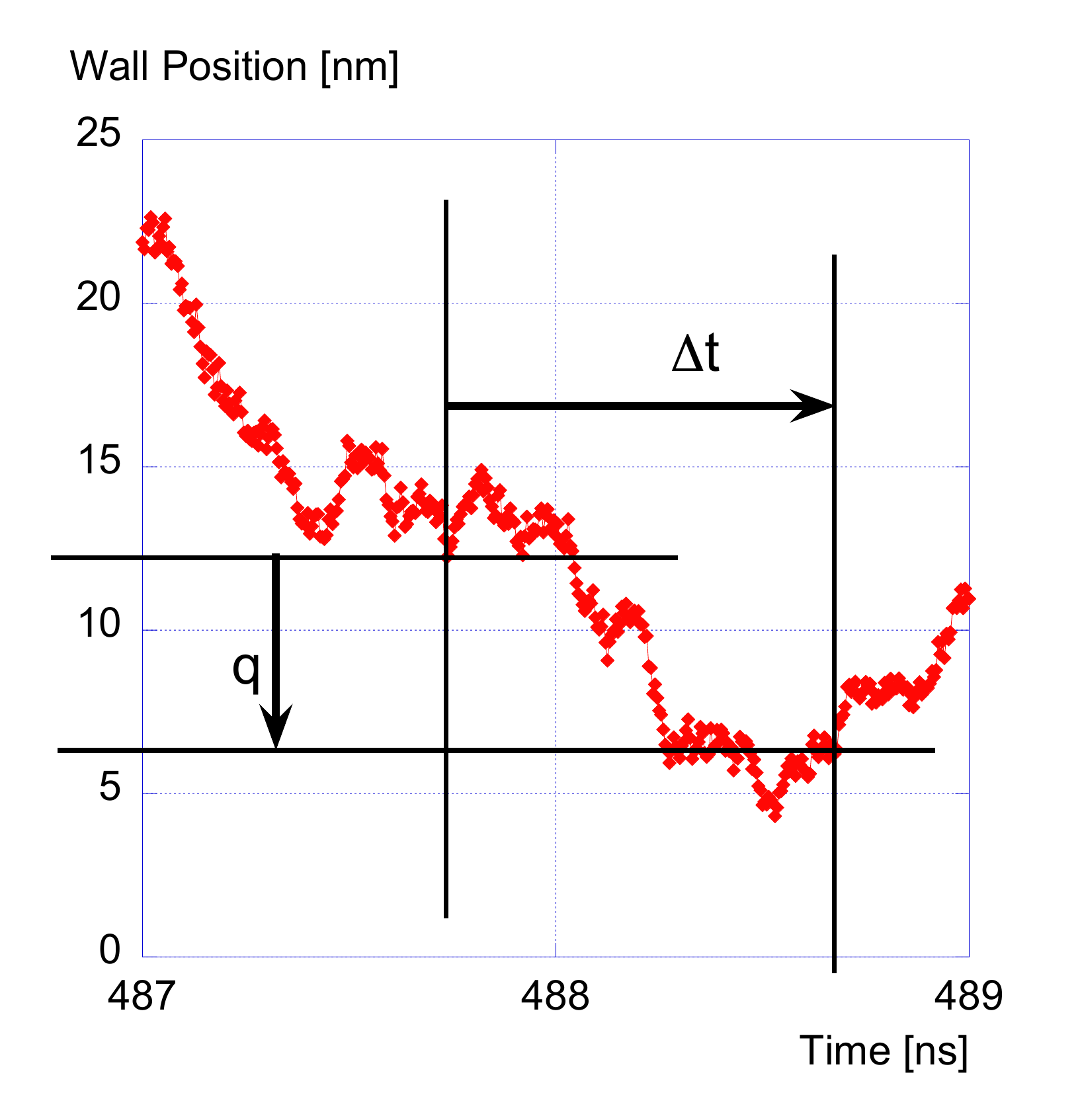}
\caption{Excerpt of a wall trace displaying wall position fluctuations \textit{vs} time: $T=77~\mathrm{K}$, $\alpha = 0.5$, $w_\mathrm{S} = 100~\textrm{nm}$, $t_\mathrm{S} = 0.6~\textrm{nm}$. $q$ is the wall displacement during time interval $\Delta t$.}
\label{Fig_WSk_Diff_2}
\end{figure}

Fig.~\ref{Fig_WSk_Diff_2} displays the position as a function of time of a wall within a $w_{\mathrm {S}} = 100~\text{nm}$ wide stripe immersed in a  $T = 77~\text{K}$ temperature bath. A $2~\text{ns}$ physical time window has been extracted from a simulation set to run for $1.5~\mu\text{s}$. The figure shows short term wall position fluctuations superimposed onto longer time diffusion. According to Einstein's theory of Brownian motion \cite{Einstein:1905}, the probability $P(x,t)$ of finding a particle at position $x$ at time $t$ obeys the classical diffusion equation $\partial_{\mathrm{t}} P(x,t) = \mathcal{D}~\partial^2_{\mathrm{x}^2} P(x,t)$ with, as a solution, a normal (gaussian) distribution $P(x,t) = 1/\sqrt{4\pi \mathcal{D}t}~\exp(-x^2 / 4\mathcal{D}t)$, where $\mathcal{D}$ is the diffusion constant. 

\begin{figure}
\centering
\includegraphics[width=0.50\textwidth]{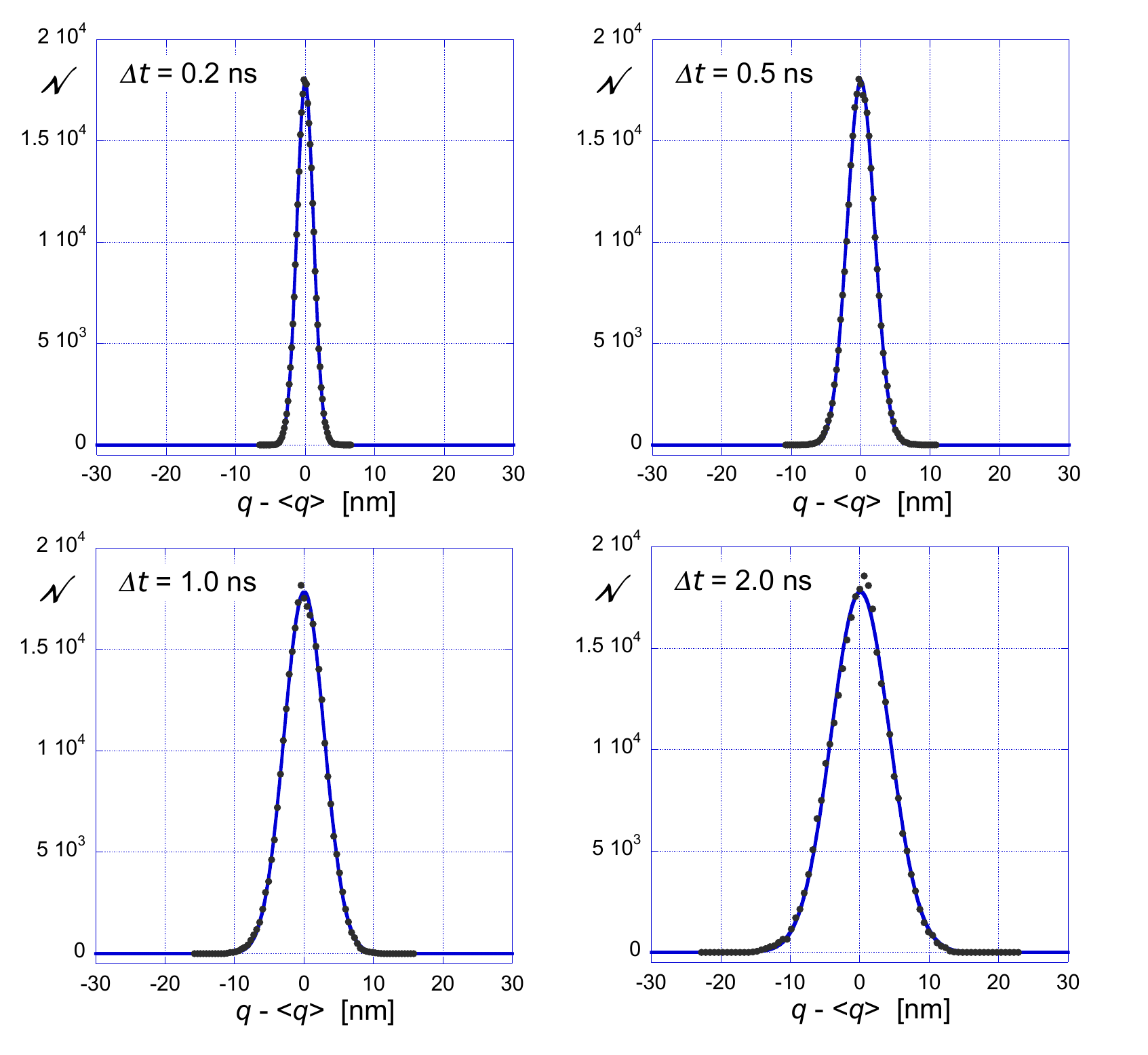}
\caption{Wall within stripe: event statistics with time interval $\Delta t$ as a parameter; $\alpha = 0.5$, $w_\mathrm{S} = 100~\mathrm{nm}$, $t_\mathrm{S} = 0.6~\textrm{nm}$, $T = 25K$. The continuous blue lines are fits to a gaussian distribution, the variance of which increases with $\Delta t$.}
\label{Fig_WSk_Diff_3}
\end{figure}

So does the raw probability of finding a (stiff) wall in a narrow stripe at position $q$ after a time interval $\Delta t$, as shown in Fig.~\ref{Fig_WSk_Diff_3} (see Fig.~\ref{Fig_WSk_Diff_2} for variable definition). It ought to be mentioned that the average wall displacement $\langle q( \Delta t ) \rangle$ is always equal to $0$, with an excellent accuracy, provided the overall computation time is large enough. The fit to a normal distribution proves rather satisfactory, with, however, as seen in Fig.~\ref{Fig_WSk_Diff_3}, a slightly increasing skewness in the distributions as a function of increasing $\Delta t$. Skewness, however, 1) remains moderate up to $\Delta t$ values typically equal to $5-10~\text{ns}$, 2) is seen to reverse sign with time interval (compare Fig.~\ref{Fig_WSk_Diff_3}b and c), excluding intrinsic biasing. The distributions standard deviation is clearly seen to increase with increasing $\Delta t$.

\begin{figure}
\centering
\includegraphics[width=0.50\textwidth]{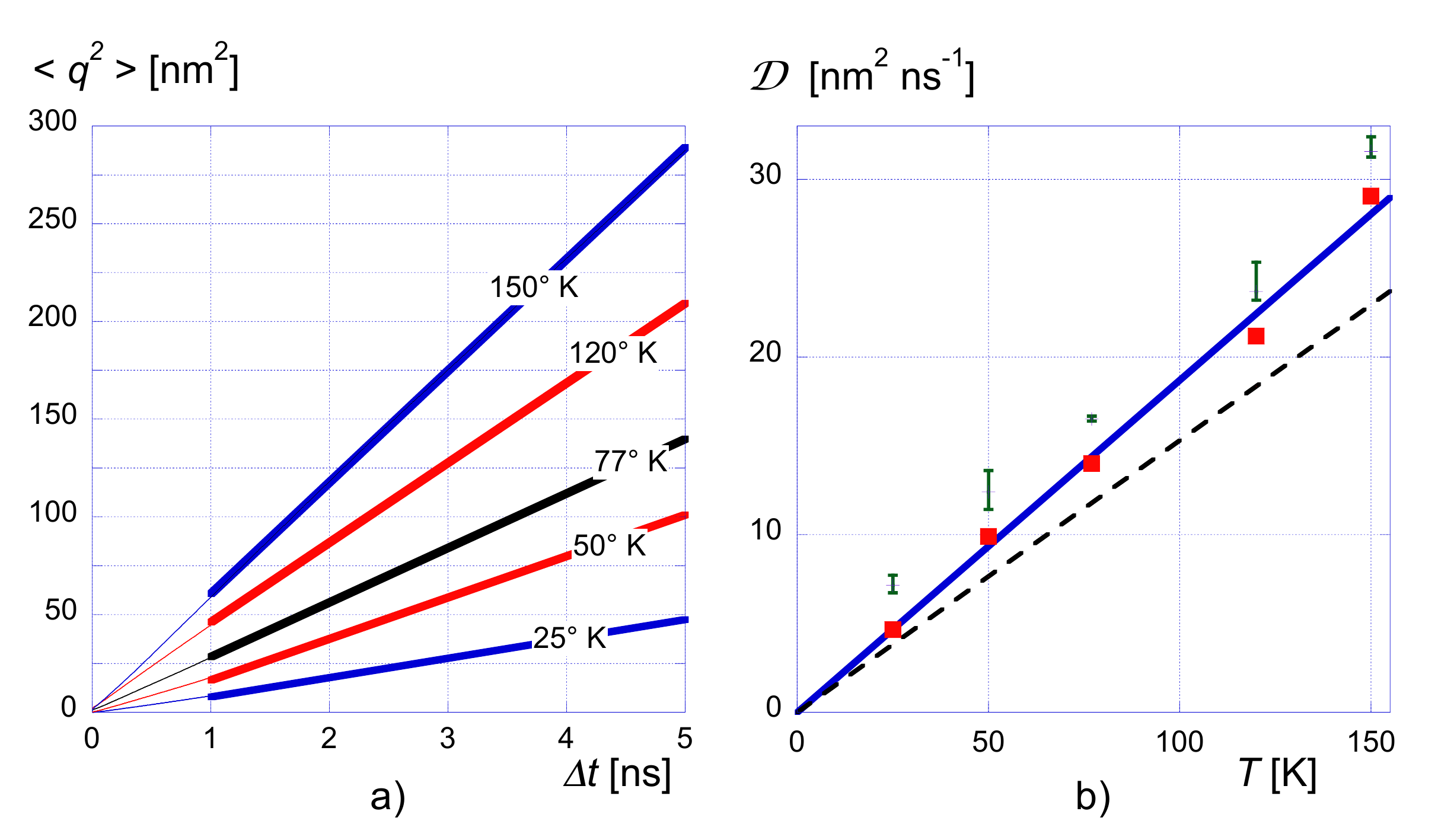}
\caption{ a) Variance $\langle q^2 \rangle$ ($\mathrm{nm}^2$) of the wall displacement \textit{vs} time interval $\Delta t$ with temperature $T$ as a parameter.  Thick lines represent a linear fit to data; 
 b) Diffusion constant $\mathcal{D}$ as a function of temperature (square full symbols). $\mathcal{D}$ is proportional to the slope of the $\langle q^2 \rangle$ \textit{vs} $\Delta t$ curves in Fig.\ref{Fig_WSk_Diff_4}a (see text for details). The error bars are deduced from the slopes of straight lines through the origin that encompass all data points in Fig.\ref{Fig_WSk_Diff_4}a for a given temperature and the fit time bracket,  $1-5~\mathrm{ns}$. For the sake of legibility, the error bars have been moved-up by $2.5$ units. Continuous line: linear fit through the origin. The dashed line is the analytical expectation in the "low" noise limit. $\alpha = 0.5$, $w_\mathrm{S} = 100~\mathrm{nm}$, $t_\mathrm{S} = 0.6~\textrm{nm}$.}
\label{Fig_WSk_Diff_4}
\end{figure}

Alternatively, one may represent the variance $\langle q^2 \rangle$ ($\langle q \rangle = 0$) as a function of the time interval $\Delta t$ : if diffusion applies, then a linear dependence is expected, with a $2 \mathcal{D}$ slope for a one-dimensional diffusion. Fig.\ref{Fig_WSk_Diff_4}a shows, for various temperatures, that a linear law is indeed observed. Lastly, as shown in Fig.\ref{Fig_WSk_Diff_4}b, the diffusion constant increases linearly with increasing temperature. The error bars measuring the departure from strict linearity in Fig.\ref{Fig_WSk_Diff_4}a remain limited in extent. For the stripe width and damping parameter considered here ($w_\mathrm{S} = 100~\text{nm}$, $\alpha = 0.5$), the ratio of diffusion constant to temperature is found to amount to $\mathcal{D} / T =  0.187~\text{nm}^2 \text{ns}^{-1} \text{K}^{-1}$.
\subsection{Wall diffusion constant (analytical)}
\label{sec:2b}
Thiele's equation \cite{Thiele:1973} states that a magnetic texture moves at constant velocity $\vec{v}$ provided the equilibrium of 3 forces be satisfied: 
\begin{equation}
\label{eq:WSkDiff:3}
\vec{G} \times \vec{v} + \mathrm \alpha \overline{\overline{D}} \vec {v} = \vec{F}
\end{equation}
where, $\vec{F}$ is the applied force, $\vec{F_G} = \vec{G} \times \vec{v}$ is the gyrotropic force, $\vec{G}$ the gyrovector, $\vec{F_D} = 
\mathrm \alpha \overline{\overline{D}} \vec {v}$ the dissipation force, $\overline{\overline{D}}$ the dissipation dyadic.

For the DMI hardened N\'eel wall considered here :  $\vec{G} = \vec{0}$. 
For a 1D wall, the Thiele equation simply reads : 
\begin{equation}
\label{eq:WSkDiff:4}
\alpha D_{xx}  v_x = F_x
\end{equation}
where, $D_{xx} =  \frac{\mu_0 M_S}{\gamma_0} \int_V (\frac{\partial{\vec{m}}}{\partial x})^2~d^3 r$.

The calculation proceeds in two steps, first evaluate the force, hence, according to Eqn.\ref{eq:WSkDiff:4},  the velocity  auto-correlation functions, then integrate \textit{vs}  time in order to derive $\langle q^2 \rangle$.
The force, per definition, is equal to minus the partial derivative of the energy $E$ w.r.t. the displacement $q$, namely $F_x = - \frac{\partial{E}}{\partial{q}} = - \mu_{0} M_{S}  \int_{V} \frac{\partial{\vec{m}}}{\partial x} \cdot \vec{H} ~d^3 r$. 
Formally, 
\begin{eqnarray}
\label{eq:WSkDiff:5}
\lefteqn{\left \langle F_x (t) F_x (t')  \right \rangle = (\mu_0 M_S)^2  \times} \\
                			 &  &    \left  \langle \int_{V} \frac{\partial \vec{m} (\vec{r} ,t)}{\partial{x}} \cdot \vec{H}(\vec{r} ,t)~d^3 r  \int_{V} \frac{\partial \vec{m} (\vec{r'},t')}{\partial{x}} \cdot \vec{H}(\vec{r'} ,t')~d^3 r' \right \rangle \nonumber
\end{eqnarray}
As noticed earlier \cite{Kamppeter:1999}, since the random field noise is "multiplicative" \cite{Brown:1963}, moving the magnetization vector out of
the average brackets is, strictly speaking, not allowed, unless considering the magnetization vector to only marginally differ from its orientation and modulus in the absence of fluctuations  (the so-called "low" noise limit \cite{Kamppeter:1999}):
\begin{eqnarray}
\label{eq:WSkDiff:6}
\lefteqn{\left \langle F_x (t) F_x (t')  \right \rangle = (\mu_0 M_S)^2 \times}  \\
                			 &  &   \int_{V}  \sum_{i,j} \left [ \frac{\partial m_i (\vec{r} ,t)}{\partial{x}} \frac{\partial m_j (\vec{r'} ,t')} {\partial{x}}
			  \left  \langle  H_i(\vec{r} ,t) H_j(\vec{r'} ,t') \right \rangle \right] d^3r \ d^3r'   \nonumber
\end{eqnarray}
If due account is being taken of the fully uncorrelated character of the thermal field (Eqn.\ref{eq.WSkDiff:1}), the force auto-correlation function becomes:
\begin{equation}
\label{eq:WSkDiff:7}
\left \langle F_x (t) F_x (t')  \right \rangle = 2 \alpha k_B T D_{xx} \delta (t-t')   
\end{equation}
The velocity auto-correlation function follows from Eqn.\ref{eq:WSkDiff:4}. Lastly, time integration ($q(t) = \int_{0}^{t} v_x(t') dt'$) yields :
\begin{equation}
\label{eq:WSkDiff:8}
\langle q^2 (t) \rangle = 2 \mathcal{D} t    \text{~~~~;~~~~}   \mathcal{D} = \frac{k_B T}{\alpha D_{xx} }    
\end{equation}
In order to relate the diffusion constant to a more directly recognizable wall mobility, $D_{xx}$ may be expanded as : 
\begin{equation}
\label{eq:WSkDiff:9}
D_{xx} = \frac{\mu_0 M_S}{\gamma_0}  \frac{2 w_\mathrm{S}  t_\mathrm{S}}{ \Delta_T }
\end{equation}
where, $\Delta_T$ has been called the Thiele wall width (implicitly defined in \cite{Thiele:1974}). $\mathcal{D}$ may thus be expressed as :
\begin{equation}
\label{eq:WSkDiff:10}
 \mathcal{D} = \frac{k_B T}{2 \mu_0 M_S} \frac{1}{w_\mathrm{S}  t_\mathrm{S}} \frac{\gamma_0 \Delta_T}{\alpha}
\end{equation}
thus, proportional to the wall mobility $\gamma_0 \Delta_T / \alpha$. 

A directly comparable result may be obtained after constructing a full Langevin equation from the ($q,\phi$) equations of domain wall motion (Slonczewski's equations \cite{Slonczewski:1972}), where $\phi$ is the azimuthal magnetization angle in the wall mid-plane. In this context, the wall mobility is $\mu_W = \gamma_0 \Delta / \alpha$, where $\Delta$ is the usual wall width, incidentally equal to the Thiele wall width in the case of a pure Bloch wall. The Langevin equation \cite{Langevin:1908} here reads:
\begin{equation}
\label{eq:JMWall-24}
\frac{m_{\mathrm D}}{2} w_{\mathrm S} t_{\mathrm S} \frac{d^2\left\langle q^2 \right\rangle}{dt^2} +
 \frac{1}{2} \frac{2 \mu_{\mathrm 0} M_{\mathrm s}}{\mu_{\mathrm W}} w_{\mathrm S} t_{\mathrm S} \frac{d\left\langle q^2\right\rangle}{dt} = 
 k_{\mathrm B} T 
\end{equation}
where, $m_{\mathrm D}$ is D\"{o}ring's wall mass density ($\mathrm{kg/m^2}$):
\begin{equation}
\label{eq:JMWall-29}
m_{\mathrm D} =  \left(1 + \alpha^2 \right) \left( \frac{\gamma_{\mathrm 0}}{2 \mu_{\mathrm 0} M_{\mathrm s}} \right)^{-2} \frac{1}{\pi \mid D_\mathrm{DM} \mid }
\end{equation}
an expression valid in the limit  $\mid D_\mathrm{DM} \mid \gg K_{\mathrm{Eff}} = K_{\mathrm{u}} - \frac{1}{2} \mu_{\mathrm 0} M_{\mathrm s}^2$. 
Note that the DMI constant $D_\mathrm{DM}$ explicitly enters the expression of the wall mass, as a consequence of the wall structure stiffening by DMI. In the stationary regime, $\langle q^2 \rangle$ is proportional to time $t$ and the wall diffusion constant exactly matches Eqn.\ref{eq:WSkDiff:10}, after substitution of $\Delta_T$ by $\Delta$. Finally, the characteristic time for the establishment of stationary motion is:
\begin{equation}
\label{eq:JMWall-30}
t_{\mathrm 0} = m_{\mathrm D} \frac{1}{2 \mu_{\mathrm 0} M_{\mathrm s}} \frac{\gamma_{\mathrm 0} \Delta}{\alpha}
\end{equation}
For the parameters of our model 3-ML Co layer on top of Pt, D\"{o}ring's mass density is equal to $\sim 3 \: 10^{-8} \mathrm{kg/m^2}$ for $ \alpha=0.5 $, and the characteristic time amounts to $ t_{\mathrm 0} \simeq 25 \:  \mathrm{ps} $.
Still for $ \alpha=0.5 $, $w_\mathrm{S}=100~\mathrm{nm}$ and  $t_\mathrm{S}=0.6~\mathrm{nm}$, $\mathcal{D} / T$ amounts to $0.153~\text{nm}^2 \text{ns}^{-1} \text{K}^{-1}$  for $\Delta_T = 4. 13~\text{nm}$, i.e. the value computed from a properly converged wall profile at $T =0$. The relative difference between simulation and theoretical values is found to be of the order of $\approx 20$\%.
\begin{figure}
\centering
\includegraphics[width=0.50\textwidth]{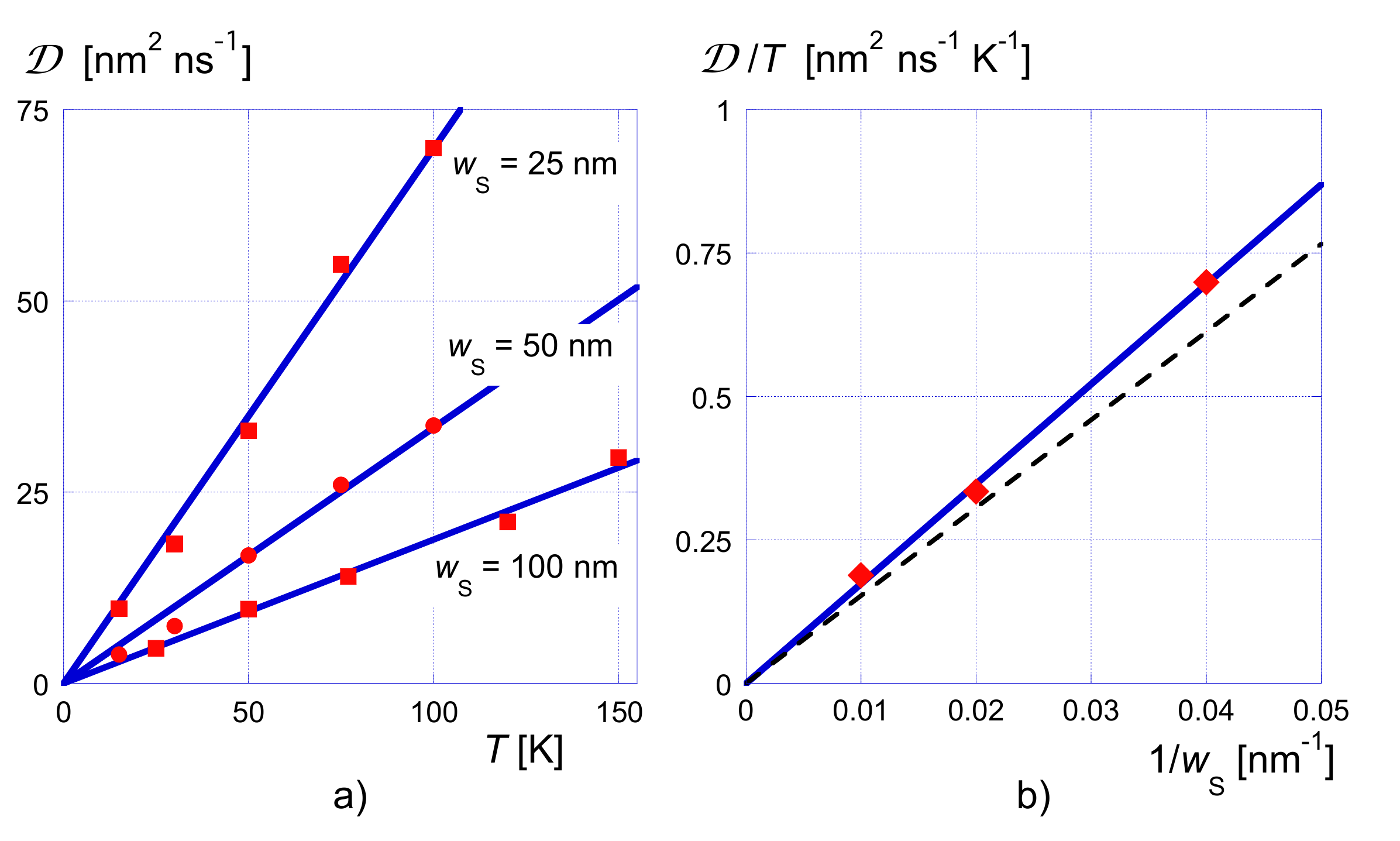}
\caption{ a) Diffusion constant $\mathcal{D}$ as a function of temperature with the stripe width $w_\mathrm{S}$ as a parameter (full symbols);  b) $\mathcal{D}/T$  as a function of the inverse of the stripe width. $\alpha = 0.5$, $t_\mathrm{S} = 0.6~\textrm{nm}$, throughout. Solid blue lines: linear fit through the origin, dashed line: analytical expectation.}
\label{Fig_WSk_Diff_5}
\end{figure}

\begin{figure}
\centering
\includegraphics[width=0.50\textwidth]{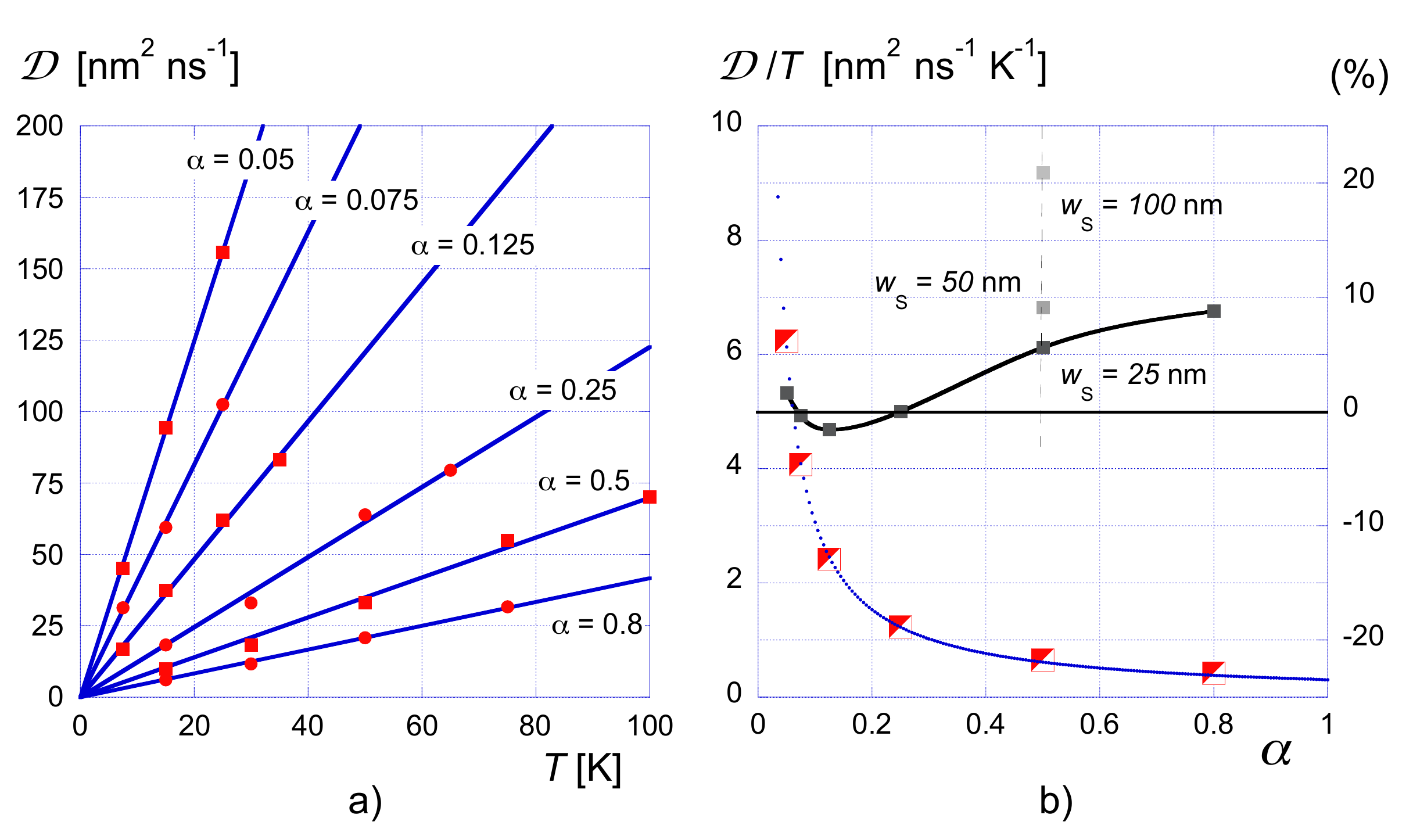}
\caption{ a) Diffusion constant $\mathcal{D}$ as a function of temperature with the damping constant $\alpha$ as a parameter ($w_\mathrm{S} = 25~\mathrm{nm}$, $t_\mathrm{S} = 0.6~\textrm{nm}$). Solid blue lines: linear fit through the origin; b) $\mathcal{D}/T$ (large semi-open symbols) as a function of $\alpha$ for $w_\mathrm{S} = 25~\mathrm{nm}$ and $t_\mathrm{S} = 0.6~\textrm{nm}$; dotted blue curve: analytical expectation. Full symbols: relative difference between computational and analytical results (\%). }
\label{Fig_WSk_Diff_6}
\end{figure}

Owing to Eqn.\ref{eq:WSkDiff:10}, $ \mathcal{D}$ is expected to prove inversely proportional to both the stripe width $w_\mathrm{S}$ and the Gilbert damping parameter $\alpha$, a behavior confirmed by simulations. Fig.\ref{Fig_WSk_Diff_5}a displays the computed values of the diffusion coefficient as a function of temperature with the stripe width as a parameter, whilst Fig.\ref{Fig_WSk_Diff_5}b states the linear behavior of 
$\mathcal{D}$ \textit{vs} ${w_{\mathrm {S}}}^{-1}$. The slope proves, however, some $13.5$\% higher than anticipated from Eqn.\ref{eq:WSkDiff:10}. Lastly, the $1/{\alpha}$ dependence is verified in Fig.\ref{Fig_WSk_Diff_6} showing the computed variation of $\mathcal{D}$ \textit{vs} temperature with $\alpha$ as a parameter for a narrow stripe (${w_{\mathrm {S}}} = 25~\text{nm}$) as well as the corresponding $\alpha$ dependence of $\mathcal{D} / T$. The dotted line represents Eqn.\ref{eq:WSkDiff:10} without any adjusting parameter. The relative difference between simulation data and theoretical expectation is beyond, say $\alpha = 0.25$, seen to grow with increasing $\alpha$ but also appears to be smaller for a narrow stripe as compared to wider tracks.

Altogether, simulation results only moderately depart from theoretical predictions. The Brownian motion of a DMI-stiffened wall in a track clearly proves diffusive. The diffusion constant is classically proportional to the wall mobility and inversely proportional to the damping parameter. Unsurprisingly, the smaller the track width, the larger the diffusion constant. In order to provide an order of magnitude, the diffusion induced displacement expectation, 
$\sqrt{2 \mathcal{D} \Delta t}$, for a wall sitting in a $100~\mathrm{nm}$-wide, pinning-free, track for 25 ns at $T = 300~\mathrm{K}$ proves essentially equal to $\pm$ the stripe width. 
\section{\label{SkDiff}Skyrmion diffusion\protect\\ }
\begin{figure}
\centering
\includegraphics[width=0.35\textwidth]{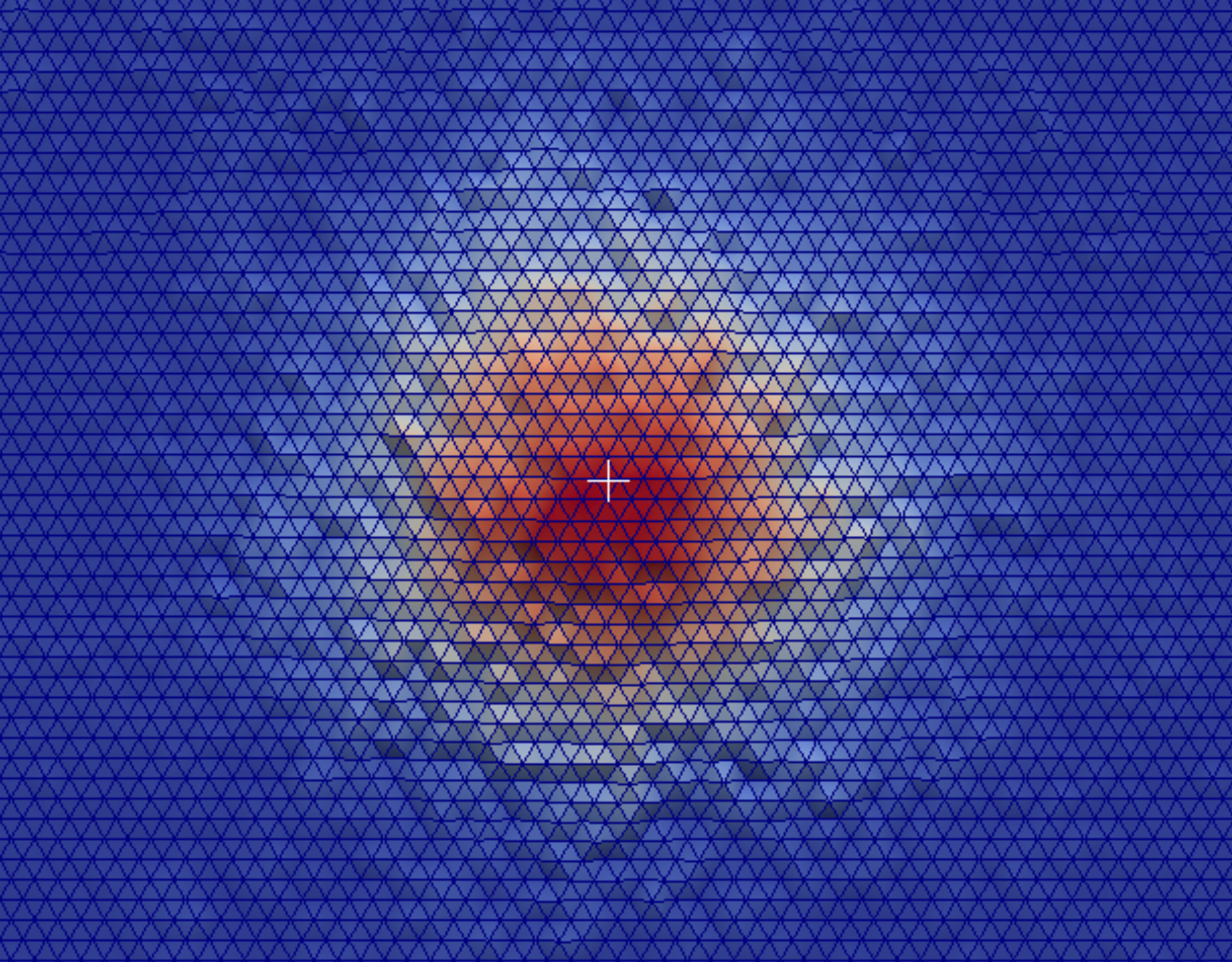}
\caption{ a) Snapshot of a skyrmion immersed in a 12.5 K temperature bath ($\alpha = 0.5$), together with the underlying lattice. Red cells: $s_z \approx +1$, blue cells:  $s_z \approx -1$. The white cross indicates the barycenter of lattice site positions satisfying $s_z \geq 0.5$. }
\label{Fig_WSk_Diff_7}
\end{figure}
Outstanding observations, by means of Spin Polarized Scanning Tunneling Microscopy, have revealed the existence of isolated, nanometer size,  skyrmions in ultra-thin films such as a PdFe bilayer on an Ir(1111) single crystal substrate \cite{Romming:2013} \cite{Romming:2014}. We analyse below the thermal motion of skyrmions in a model system made of a Co ML on top of Pt(111). We deal with skyrmions with a diameter of about $2.5~\text{nm}$ containing at $T=0$ about $250$ spins.
\subsection{\label{Sk_Simul} Simulation results}
In order to monitor the Brownian motion of an isolated skyrmion, rather than micromagnetics, it is preferred to simulate the thermal agitation of classical spins, $\vec{s}$ ($\mid s\mid = 1$),  on a triangular lattice. Lattice effects and frequency cutoffs in thermal excitations are thus avoided. Such simulations have already been used e.g. for the determination of the barrier to collapse of an isolated skyrmion \cite{Rohart:2016,Rohart:2017}.
The parameters are: lattice constant $a = 2.51$~\AA, magnetic moment $\mu_{At}=2.1$~$\mu_B$/atom, Heisenberg exchange nearest neighbor constant $J=29$~meV/bond, Dzyaloshinskii-Moriya exchange $d=-1.5$~meV/bond, magnetocrystalline anisotropy $0.4~\text{meV/atom}$. The stochastic field is still defined by Eqn.\ref{eq.WSkDiff:1} after substitution of the product $M_S V$ by the magnetic moment per atom. The code features full magnetostatic (dipole-dipole) interactions. Fast Fourier Transforms implementation ensues from the decomposition of the triangular lattice into two rectangular sublattices, at the expense of a multiplication of the number of dipole-dipole interaction coefficients. Lastly, the base time step, also the stochastic field refresh time, has been given a low value in view of the small atomic volume, namely $dt = 2.5~\text{fs}$ for $\alpha \geq 0.1$, $dt = 1~\text{fs}$ below. Time steps that small may be deemed little compatible with the white thermal noise hypothesis \cite{Brown:1963}. They are in fact dictated by the requirement for numerical stability, primarily w.r.t. exchange interactions. 

\begin{figure}
\centering
\includegraphics[width=0.50 \textwidth]{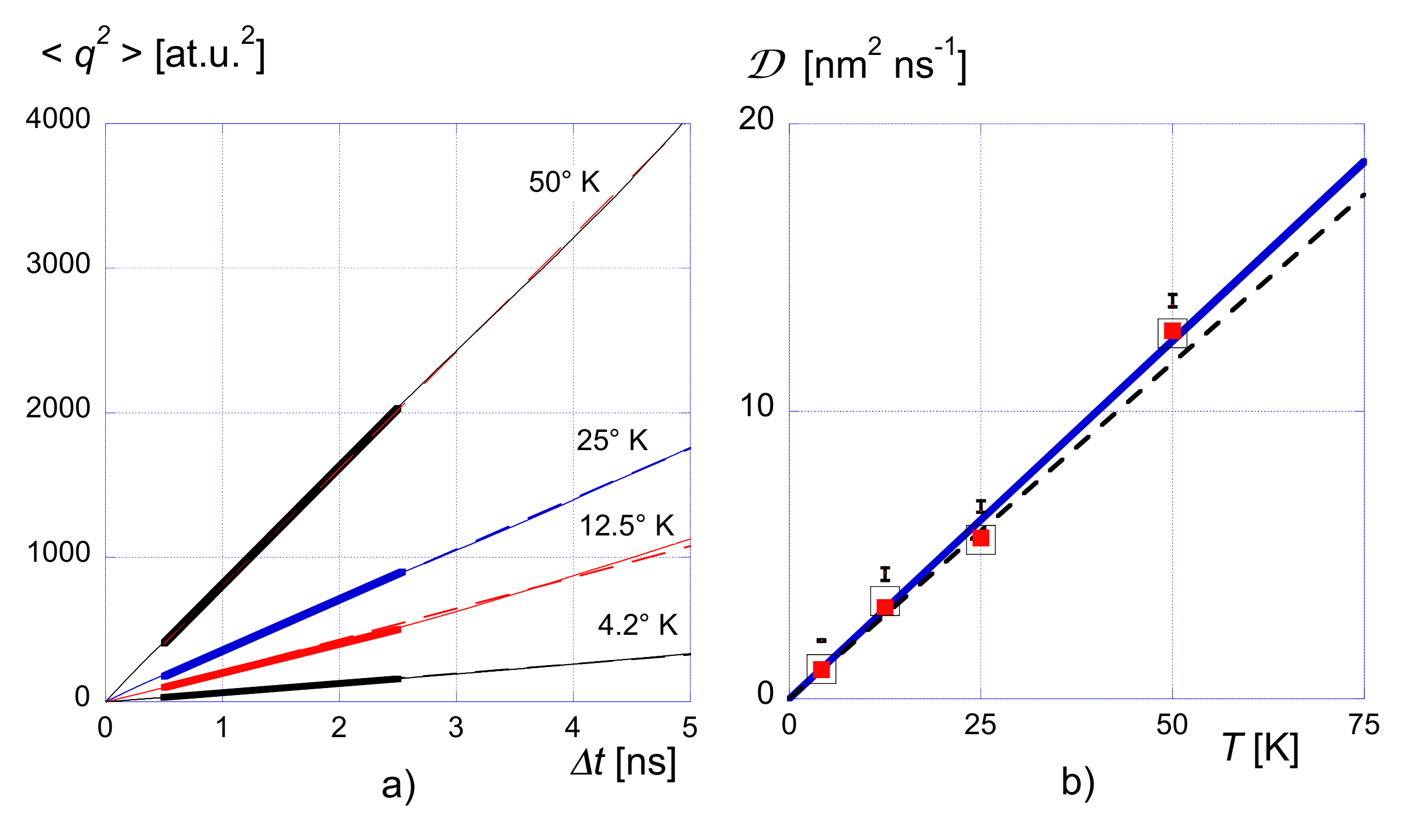}
\caption{
a) Variance ($\mathrm{at.u.}^2$) of the skyrmion displacement $\langle q^2 \rangle$ \textit{vs} time interval $\Delta t$ with temperature $T$ as a parameter.  Thick and dashed lines represent a linear fit to data with different time coverage, namely $[0.25 - 2.5~\mathrm{ns}]$ and $[0 - 5~\mathrm{ns}]$; 
 b) Diffusion constant $\mathcal{D}$ as a function of temperature for a $[0.25-2.5~\mathrm{ns}]$- (open symbols) and $[0-5~\mathrm{ns}]$- (full symbols) linear fit. Solid blue line: linear fit through the origin. Dashed line: analytical expectation in the "low" noise limit. In order to ensure legibility, the error bars as defined in the caption of Fig.\ref{Fig_WSk_Diff_4} and pertaining to the $[0.25-2.5~\mathrm{ns}]$ fit time bracket have been moved-up by one unit. $\alpha = 0.5$.}
\label{Fig_WSk_Diff_10} 
\end{figure}

Fig.\ref{Fig_WSk_Diff_7} is a snapshot of an isolated skyrmion in the model Co ML with a temperature raised to $12.5~\text{K}$. 
The skyrmion is at the center of a $200~\text{at.~u.}$- i.e. $\approx 50~\text{nm}$-size square computation window, that contains 46400 spins and is allowed to move with the diffusing skyrmion. Doing so alleviates the computation load without restricting the path followed by the skyrmion. Free boundary conditions (BC's) apply. The window, however, proves sufficiently large to render the confining potential created by BC's ineffective. 
\begin{figure}
\centering
\includegraphics[width=0.40 \textwidth]{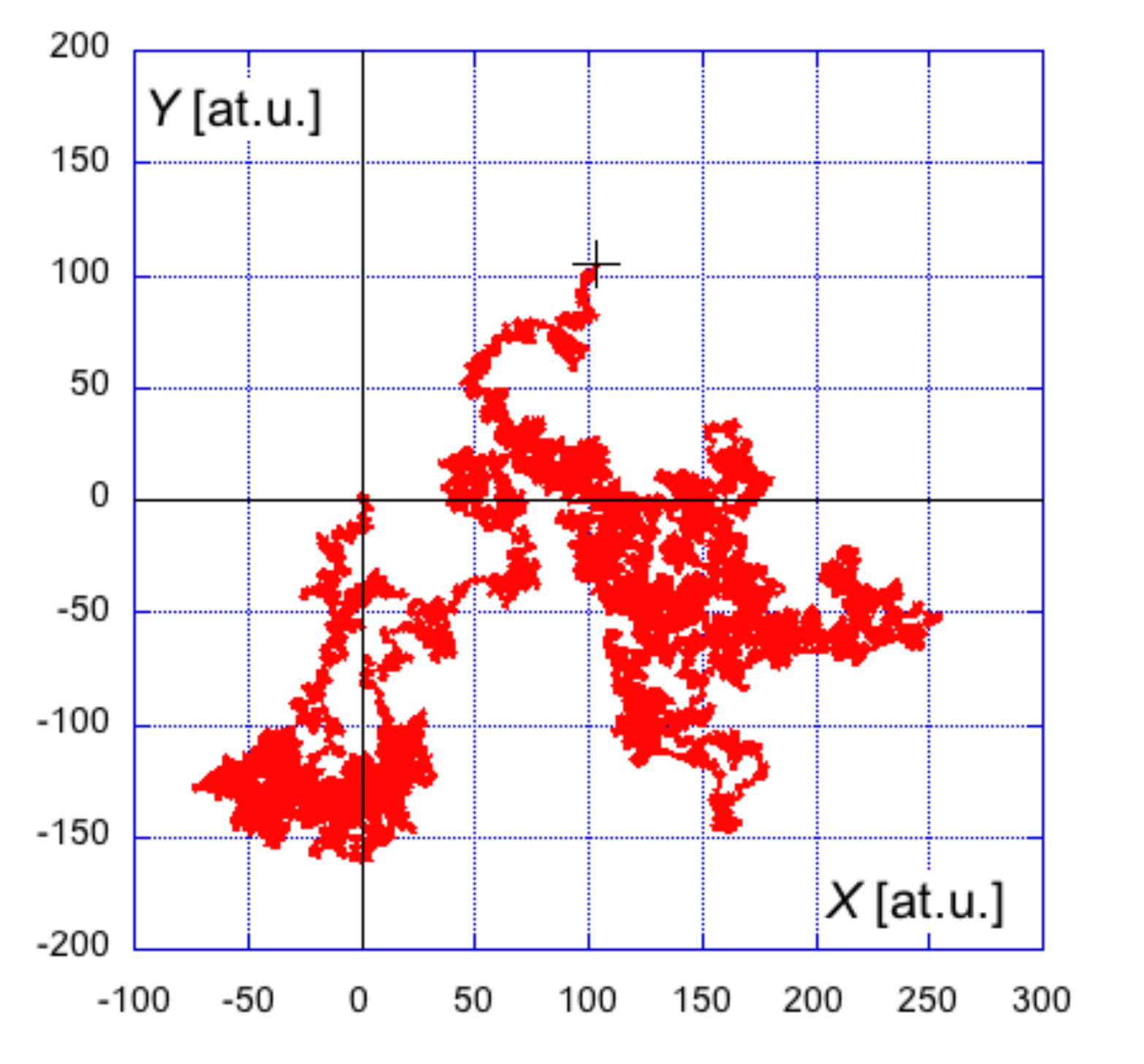}
\caption{ Example of skyrmion trajectory. Distances in atomic units ($1~\mathrm{at.u.} = 2.51~\mathrm{\AA}$).  The trajectory started at the origin of  coordinates at time $t=0$ and stopped at the cross location at physical time $t \approx{100}~\text{ns}$. $T=25~\mathrm{K}$, $\alpha =1$. }
\label{Fig_WSk_Diff_8}
\end{figure}
The skyrmion position as a function of time is defined simply as the (iso)barycenter of the contiguous lattice site positions $x(k)$, $y(k)$, where $s_z \geq 0.5$:
\begin{equation}
\label{eq:WSkDiff:11}
q_x^{Sk} = \frac{1}{K} \sum_{k=1}^{K} x(k) \text{      ; } q_y^{Sk} = \frac{1}{K} \sum_{k=1}^{K} y(k)
\end{equation}
where, $k$ is the lattice site index, $K$ the number of lattice sites satisfying the above condition. Such a definition proves robust \textit{vs}  thermal disorder such as displayed in Fig. \ref{Fig_WSk_Diff_7}. Similarly to the case of wall diffusion, we analyze first the distributions of the displacement components $q_x, q_y$.  
\begin{figure}
\centering
\includegraphics[width=0.50 \textwidth]{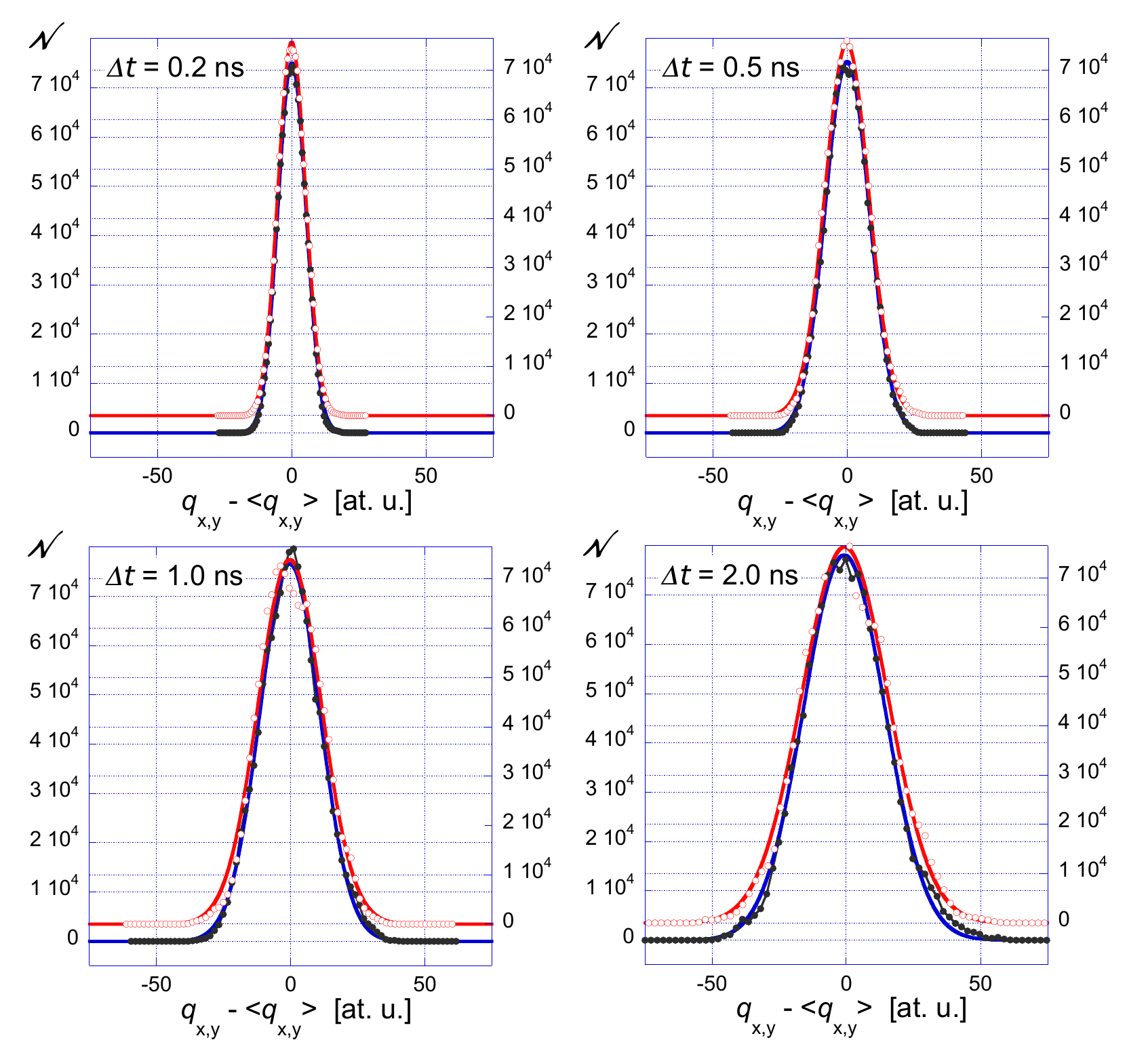}
\caption{ Skyrmion: event statistics with time interval $\Delta t$ as a parameter for the displacement components $q_x$ (black full symbols) and $q_y$ (red open symbols), labeled $q_{x,y}$ in the figures. In each panel, the curves have been offset vertically for legibility. Solid lines: fit to a gaussian distribution. $\alpha =  0.25$, $T = 25~\textrm{K}$}
\label{Fig_WSk_Diff_9}
\end{figure}
The event statistics for each value of the time interval is clearly gaussian (see Fig.\ref{Fig_WSk_Diff_9}). However, the noise in the distributions appears larger when compared to the wall case. It also increases faster with $\Delta t$. On the other hand, the raw probabilities for $\langle q_x^2 \rangle$ and $\langle q_y^2 \rangle$ barely differ as anticipated from a random process. The behavior of $\langle q^2 \rangle$  
($q^2 = q_x^2+q_y^2$) \textit{vs} $\Delta t$ is displayed in Fig.\ref{Fig_WSk_Diff_10}a. 

The range of accessible temperatures is governed by the thermal stability of the tiny skyrmion within a Co ML: with a lifetime of $ \simeq 1~\mathrm{\mu s}$ at $77~\text{K}$ \cite{Rohart:2016,Lobanov:2016, Bessarab:2017,Rohart:2017}, temperatures have been confined to a $\leq 50~\mathrm{K}$ range. When compared to the wall case (Fig.\ref{Fig_WSk_Diff_4}a), the linear dependence of  $\langle q^2 \rangle$ with respect to $\Delta t$ appears less satisfactory, although, over all cases examined, the curves do not display a single curvature, but rather meander gently around a straight line. The slope is defined as the slope of the linear regression either for time intervals between $0.25$ and $2.5$ ns (thick line segments in Fig.\ref{Fig_WSk_Diff_10}a) or for the full range $0$ to $5$ ns (dashed lines). Then, the ratio of the diffusion constant to temperature, $\mathcal{D} / T$, for an isolated skyrmion within the model Co ML considered here is equal to $0.250$ and $0.249~\text{nm}^2 \text{ns}^{-1} \text{K}^{-1}$, respectively, for $\alpha = 0.5$ (see Fig.\ref{Fig_WSk_Diff_10}b). The difference proves marginal. Lastly, error bars appear even narrower than in the wall case. 
\subsection{Skyrmion diffusion constant (analytical)}
\label{sec:3b}
The gyrovector $\vec G$ in Thiele's equation (Eqn.\ref{eq:WSkDiff:3}) has in the case of a skyrmion or a vortex, and in many other instances such as lines within walls, a single non-zero component, here $G_z$. Thiele's equation, in components form, reads:
\begin{equation}
\label{eq:WSkDiff:12}
\begin{split}
- G_\mathrm{z} v_\mathrm{y} + \alpha \left[ D_\mathrm{xx} v_\mathrm{x} + D_\mathrm{xy} v_\mathrm{y} \right] = F_\mathrm{x}
\\
+ G_\mathrm{z} v_\mathrm{x} + \alpha \left[ D_\mathrm{yx} v_\mathrm{x} + D_\mathrm{yy} v_\mathrm{y} \right] = F_\mathrm{y}
\end{split}
\end{equation}
Because of the revolution symmetry of a skyrmion at rest, $D_{xy}$ or $D_{yx}$ may safely be neglected and $D_{yy} = D_{xx} $ . Accordingly, the velocities may be expressed as: 
\begin{equation}
\label{eq:WSkDiff:13}
v_x = \frac{\alpha D F_x + G F_y}{G^2 + (\alpha D)^2} \text{    ; } v_y = \frac{\alpha D F_y - G F_x}{G^2 + (\alpha D)^2}
\end{equation}
where, $G = G_z$, $D = D_{xx} = D_{yy}$.

Similarly to the stochastic field, the force components are necessarily uncorrelated. The velocity autocorrelation functions may now be obtained following the same lines as in the wall case, yielding, in the low noise approximation:
\begin{equation}
\label{eq:WSkDiff:14}
\left \langle v_x (t) v_x (t')  \right \rangle=\left \langle v_y (t) v_y (t')  \right \rangle = 2 k_B T \frac{\alpha D}{G^2+(\alpha D)^2} \delta (t-t')
\end{equation}
The average values of the displacements squared, $\langle q_x^2 \rangle$ and $\langle q_y^2 \rangle$ follow from time integration:
\begin{equation}
\label{eq:WSkDiff:15}
\left \langle q_x^2(t) \right \rangle = \left \langle q_y^2(t) \right \rangle = 2 k_B T \frac{\alpha D}{G^2+(\alpha D)^2}~t
\end{equation}
As shown previously \cite{Schutte:2014, Troncoso:2014}, the diffusion constant for a skyrmion thus reads:  
\begin{equation}
\label{eq:WSkDiff:16}
\mathcal D =  k_B T \frac{\alpha D}{G^2+(\alpha D)^2}
\end{equation}
\begin{figure}
\centering
\includegraphics[width=0.40 \textwidth]{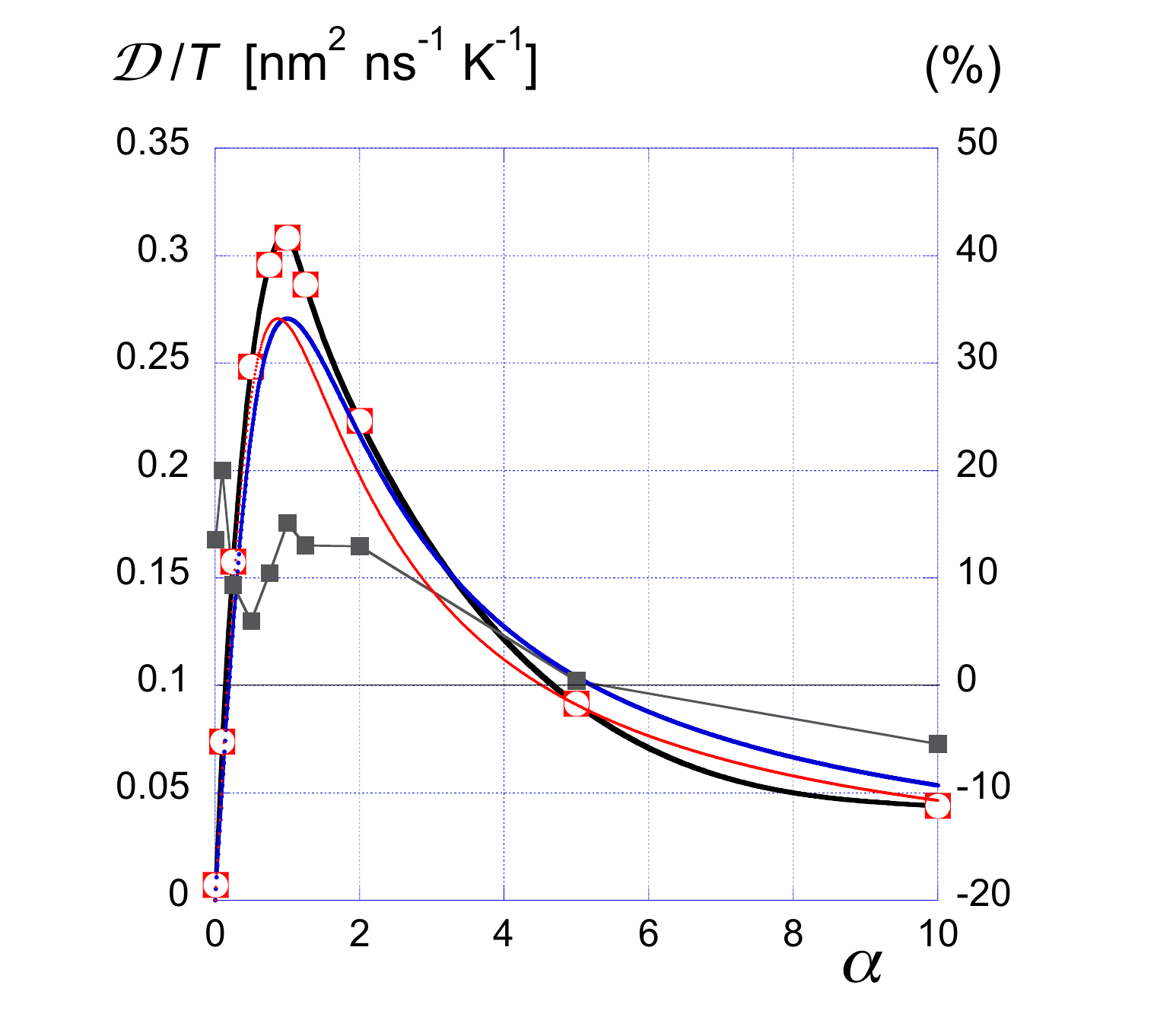}
\caption{ 
Computed values of $\mathcal{D} /T$ \textit{vs} $\alpha$ (large open symbols); black line: guide to the eye; blue (resp. red) solid curves: analytical values with $[\gamma_0 S_{At}/\mu_0 \mu_{At}] D = 4 \pi$ (resp. $14.5$). The blue curve thus corresponds to the Belavin-Polyakov profile limit. 
The relative difference between simulation and theory is indicated by small full symbols (\% : right scale).}
\label{Fig_WSk_Diff_11} 
\end{figure}
The following relations do apply:
\begin{equation}
\label{eq:WSkDiff:17}
\begin{split}
\left \langle q_x^2(t) \right \rangle = \left \langle q_y^2(t) \right \rangle = 2 \mathcal D t 
\\
\left \langle q^2(t) \right \rangle = \left \langle q_x^2(t) + q_y^2(t) \right \rangle = 4 \mathcal D t 
\end{split}
\end{equation}

Relation (\ref{eq:WSkDiff:16}) implies a peculiar damping constant dependence with, assuming for the time being $D$ and $G$ to have comparable values, a gradual drop to zero of the diffusion constant with decreasing $\alpha$ ($\alpha \le 1$), termed "diffusion suppression by $G$" by C. Sch{\"u}tte \textit{et al.} \cite{Schutte:2014}. 
Diffusion suppression is actually not a complete surprise since, for electrons in a magnetic field, a similar effect is leading to the classical magnetoresistance. 
A similar dependence 
$\mathcal D (\alpha)$ is also expected for a vortex. Boundary conditions, however, add complexity to vortex diffusion. What nevertheless remains, is a linear dependence of $\mathcal D$ \textit{vs} $\alpha$  \cite{Kamppeter:1999}, namely, diffusion suppression. 

The classical expressions for $G_z$ and $D_{xx}$ valid for a magnetization continuum need to be adapted when dealing with discrete spins. We obtain: 
\begin{equation}
\label{eq:WSkDiff:18}
\begin{split}
G_z &= \frac{\mu_0  \mu_{At}}{\gamma_0}  \sum_k
\left[  \vec{s}(k) \cdot  \left[ \partial_x  \vec{s}(k) \times  \partial_y \vec{s}(k) \right] \right] 
\\
D_{xx} &=  \frac{\mu_0  \mu_{At}}{\gamma_0} \sum_k \left[
\left[ \partial_x \vec{s}(k) \right]^2 \right]  
\end{split}
\end{equation}
where, $\mu_{At}$ is the moment per atom. 

The dimensionless product $\frac{\gamma_0 S_{At}}{\mu_0  \mu_{At}} G_z$ (Eqn.\ref{eq:WSkDiff:18}), where $S_{At}$ is the surface per atom, amounts to $4 \pi$, irrespective of the skyrmion size in a perfect material at $T=0$. Stated otherwise, the skyrmion number is 1 \cite{Nagaosa:2013}. In the Belavin-Polyakov profile limit \cite{Belavin1975}, the dimentionless product $\frac{\gamma_0 S_{At}}{\mu_0  \mu_{At}} D_{xx}$ (Eqn.\ref{eq:WSkDiff:18}) also amounts to $4 \pi$. In this limit, $\mathcal D$ is proportional to $\alpha / (1+\alpha^2)$.
$D_{xx}$ increases with skyrmion radius beyond the Belavin-Polyakov profile limit (see supplementary material in \cite{Hrabec:2017}). For a skyrmion at rest in the model Co ML considered here, $D = D_{xx} \approx 14.5~\mu_0 \mu_{At} / (\gamma_0 S_{At})$. For that value of $D_{xx}$, and for the parameters used in the simulations, $\mathcal{D} / T$, the ratio of the theoretical skyrmion diffusion constant to temperature, is equal $0.234~\text{nm}^2 \text{ns}^{-1} \text{K}^{-1}$, for $\alpha = 0.5$ ($S_{At} = a^2 \sqrt{3}/2$),  to be compared to the $0.250$ value extracted from simulations. More generally, Fig.\ref{Fig_WSk_Diff_11} compares numerical  $\mathcal{D} / T$ values calculated for a broad spectrum of $\alpha$ values with theoretical expectations for $D = 14.5~\mu_0 \mu_{At} / (\gamma_0 S_{At})$ and in the Belavin-Poliakov limit. The average difference between analytical and simulation results is, in the $\alpha = (0,1)$ interval, seen to be of the order of $\simeq 15 \%$. 
\section{\label{Discussion}Discussion\protect\\ }
In the present study of thermal diffusion characteristics, satisfactory agreement between simulations and theory has been attained for DMI stiffened magnetic textures, be it walls in narrow tracks or skyrmions. The $\alpha$ dependence of the diffusion constants has been thoroughly investigated, with, as a result, a confirmation of Brownian motion suppression in the presence of a non-zero gyrovector or, equivalently, a topological signature. The theory starts with the Thiele relation applying to a texture moving under rigid translation at constant velocity. Furthermore, the chosen values of the components of the dissipation dyadic, are those valid for textures at rest, at $T=0$. The $\alpha$ dependence of the diffusion constants clearly survives these approximations. And, yet, a wall within a narrow stripe or a skyrmion in an ultra-thin magnetic layer are deformable textures, as obvious from Figs.\ref{Fig_WSk_Diff_1},\ref{Fig_WSk_Diff_7}. Simulations, on the other hand, rely on the pioneering analysis of Brownian motion, here meaning magnetization/spin orientation fluctuations \cite{Brown:1963}, within a particle small enough to prove uniformly magnetized and then extend the analysis to ultra-small computation cell volumes down to the single spin. Both approaches rely on the hypothesis of a white -uncorrelated- noise at finite temperature. 

The discussion of results is organized in two parts. In the first, results are analyzed in terms of a sole action of structure plasticity on the diagonal elements of the dissipation dyadic. In the second, we envisage, without further justification, how the present results are amended if, in the diffusion constants of walls and skyrmions (Eqns.\ref{eq:WSkDiff:8} and \ref{eq:WSkDiff:16}), the gyrotropic and dissipation terms are replaced by their time average as deduced from simulations. 

\subsection{Size effects}
\label{sec:4a}

\begin{figure}
\centering
\includegraphics[width=0.50 \textwidth]{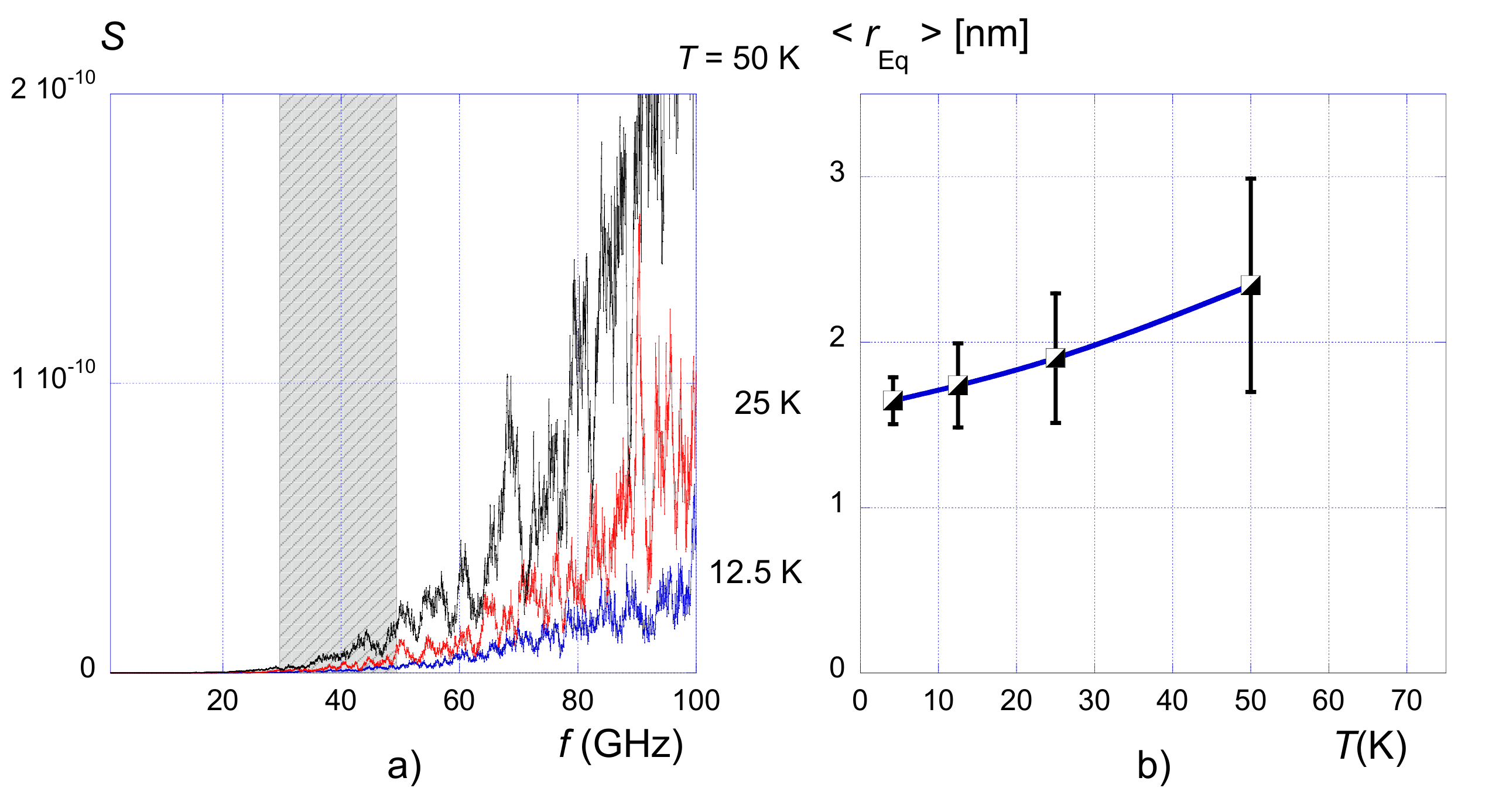}
\caption{ a) Power spectrum $S$ of the time series $r_{Eq}(t)$ for three temperatures. The hatched area corresponds to the frequency range where a signature of the fundamental skyrmion breathing mode is anticipated to be observed ($\approx 39.3~\mathrm{GHz}$, in the present case); b)  Equivalent skyrmion radius  $\langle r_{Eq} \rangle$ as a function of temperature. Error bars correspond to $\pm 1 \sigma$  of the gaussian distribution, itself a function of temperature. $\alpha = 0.5$, throughout.}
\label{Fig_WSk_Diff_12} 
\end{figure}
The integral definition of wall position adopted in this work (Eqn.\ref{eq:WSkDiff:2}) allows for a 1D treatment of wall diffusion, thus ignoring any diffusion characteristics potentially associated with wall swelling, tilting, curving or meandering. Additional information is, however, available in the case of skyrmions. We concentrate here on the number, $n$, of spins within the skyrmion satisfying the condition $s_z \geq 0.5$, and its fluctuations as a function of time. The surface of the skyrmion is $n S_{At}$ and its equivalent radius,  $r_{Eq}$, is defined by $r_{Eq}^2=nS_{At}/\pi$. The skyrmion radius $r_{Eq}$ is found to fluctuate with time around its average value, according to a gaussian distribution that depends on temperature, but becomes independent of the autocorrelation time interval beyond $\approx 25~\mathrm{ps}$. The power spectrum of the time series $r_{Eq}(t)$, shown in Fig.\ref{Fig_WSk_Diff_12}a, excludes the existence of a significant power surge around the fundamental breathing mode frequency of the skyrmion ($\approx 39.3~\mathrm{GHz}$ for the present model Co ML) \cite{Kim:2014}. The skyrmion radius as defined from the discrete $n$ distribution is thus subject to white noise. The average radius $\langle r_{Eq} \rangle$, on the other hand, varies significantly with temperature, increasing from $\approx 1.6~\textrm{nm}$ to $2.4~\textrm{nm}$ when the temperature is increased from $4.2~\textrm{K}$ to $50~\textrm{K}$ (Fig.\ref{Fig_WSk_Diff_12}b) and the diagonal element of the dissipation dyadic is expected to increase with increasing skyrmion radius \cite{Sampaio:2010, Hrabec:2017}.

Owing to relations (\ref{eq:WSkDiff:16},\ref{eq:WSkDiff:18}), the maximum of $\mathcal D (\alpha)$ is found for $\alpha=G_z/D_{xx}=G/D$. For $\alpha < G/D$, resp. $\alpha > G/D$, $\mathcal D$ increases, resp. decreases, with $D$, hence the relative positions of the blue and black continuous curves in Fig.\ref{Fig_WSk_Diff_11}.  At maximum, $\mathcal D$ is independent of $D$ and amounts to $  k_B T \frac{\gamma_0 S_{At}}{\mu_0 \mu_{At}} \frac{1}{2G} = k_B T \frac{\gamma_0 S_{At}}{\mu_0 \mu_{At}} \frac{1}{8 \pi}$. It ensues that the discrepancy between numerical and analytical $\mathcal D$ values around $\alpha = 1$ may not be relaxed by a sole variation of $D$. On the other hand, allowing $D$ to increase with skyrmion radius, itself a function of temperature, leads to an increase (decrease) of the diffusion coefficient for $\alpha < G/D$ ($\alpha > G/D$).

Likely more important is the reduction, as a function of skyrmion size, of the $\alpha$ window where diffusion suppression is expected.  
If including the $(R/\Delta+\Delta/R)$ dependence of $D_{xx}$ (see supplementary material in \cite{Hrabec:2017}; $\Delta$ is the wall width and $R$ the skyrmion radius), the skyrmion diffusion constant may be expressed as:
\begin{equation}
\label{eq:WSkDiff:19}
\begin{split}
\mathcal{D} &=  k_B T \frac{\gamma_0 S_{At}}{\mu_0 \mu_{At}} \frac{1}{8 \pi} f \left(\alpha, \frac{R}{\Delta}\right)
\\
\eta &= \frac{R}{\Delta} \textrm{    ; } \xi = \frac{1}{2} \left( \frac{1+\eta^2}{\eta} \right)  \textrm{    ; } f(\alpha, \eta) = \frac{2 \alpha \xi}{1+(\alpha \xi)^2}
\end{split}
\end{equation}
\begin{figure}
\centering
\includegraphics[width=0.45 \textwidth]{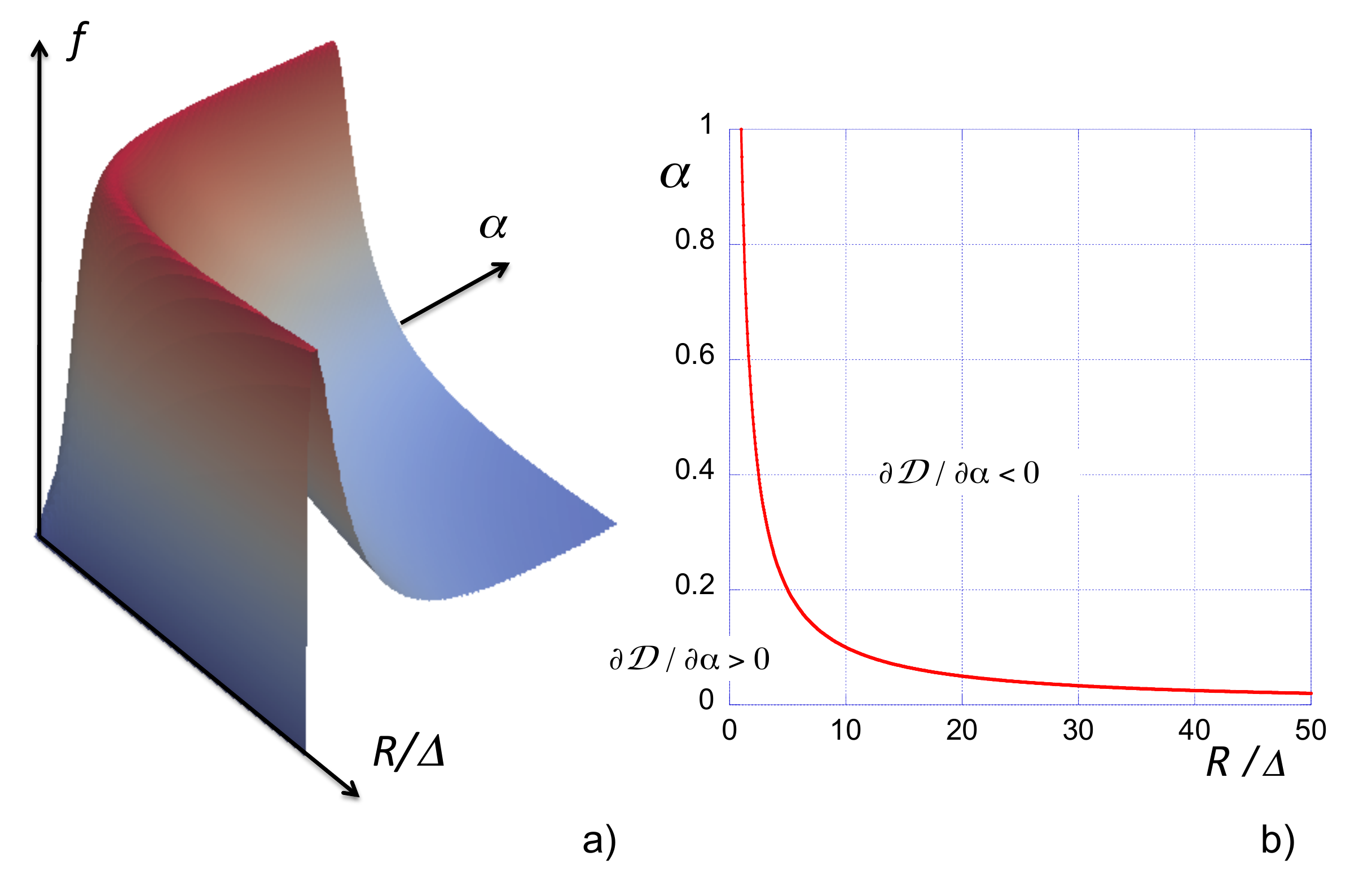}
\caption{Diffusion suppression: a) general shape of function $f(\alpha, R/\Delta)$ with $ 0 < \alpha < 1$, $ 1 < R/\Delta < 50$; b) crest line separating the region of diffusion suppression ($\partial \mathcal{D}/\partial \alpha > 0$) from region $\partial \mathcal{D}/\partial \alpha < 0$.}
\label{Fig_WSk_Diff_13} 
\end{figure}
The general shape of function $f(\alpha,R/\Delta)$ is shown in Fig.\ref{Fig_WSk_Diff_13}a. The maximum of $f(\alpha,R/\Delta)$ is equal to $1$ for all values of $\alpha$ and $R/\Delta$. The crest line $R \alpha=\Delta$ is seen to divide the parameter space into two regions (see Fig.\ref{Fig_WSk_Diff_13}b), a region close to the axes where $\partial \mathcal{D}/\partial \alpha > 0$, i.e. the region of diffusion suppression, from the much wider region where $\partial \mathcal{D}/\partial \alpha < 0$, that is, the region of wall-like behavior for skyrmion diffusion. Clearly, the $\alpha$ window for diffusion suppression decreases dramatically with increasing skyrmion size $R/\Delta$. A first observation of skyrmion Brownian motion at a video recording time scale ($25~\textrm{ms}$)  may be found in the Supplementary Material of Ref.\cite{Jiang:2015}. Skyrmions are here unusually large and most likely escape the diffusion suppression window ($\alpha < 0.02$ for $R/\Delta = 50$). Combining skyrmion thermal stability with general observability and damping parameter tailoring may, as a matter of fact, well prove extremely challenging for the observation of topology related diffusion suppression. 
\subsection{Time averaging}
\label{sec:4b}

\begin{figure}
\centering
\includegraphics[width=0.45 \textwidth]{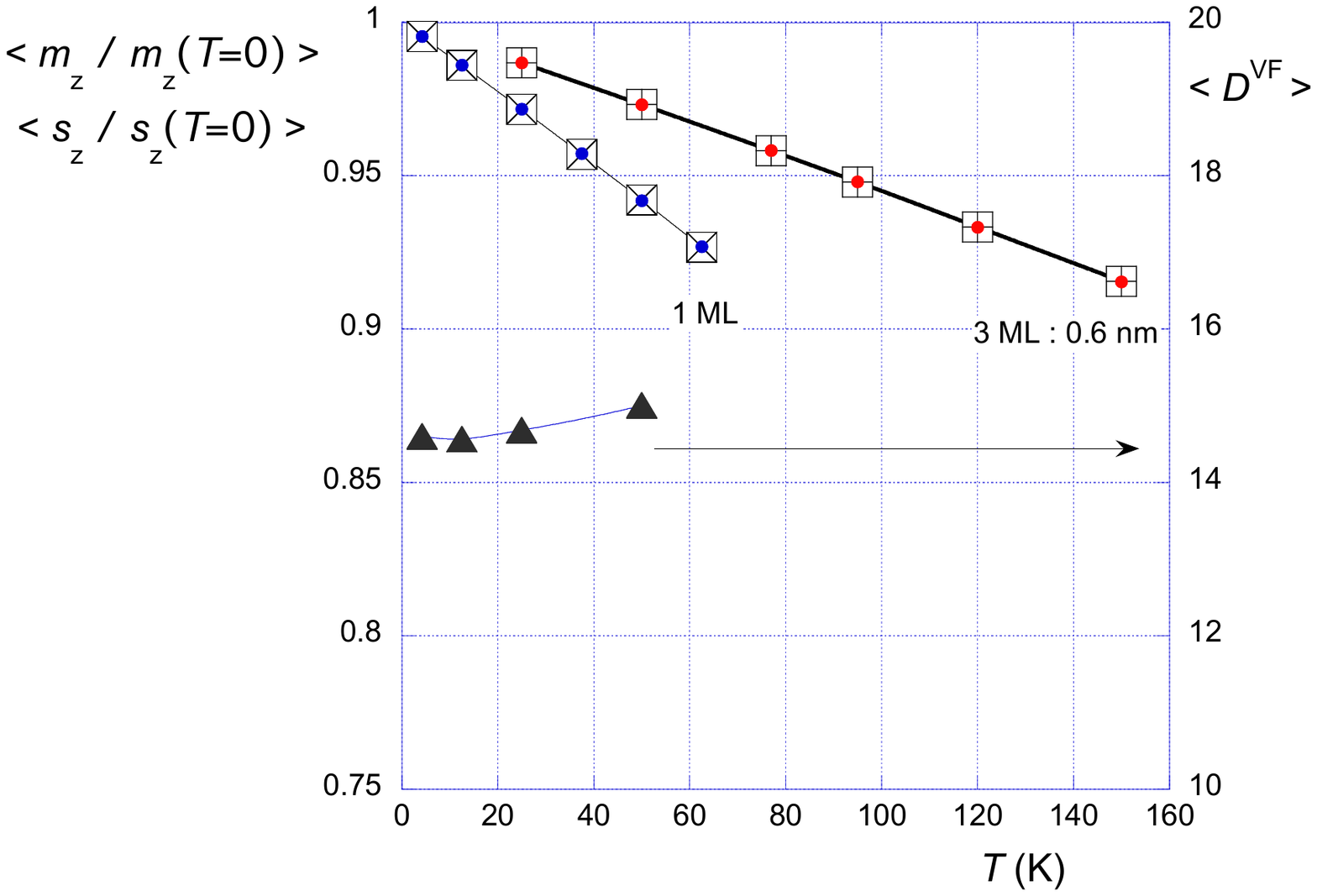}
\caption{Average reduced $z$ magnetization or spin component as a function of temperature (left scale) and time averaged value of the sole vector function, $\langle D^{\mathrm{VF}} \rangle$, within the diagonal element of the dissipation tensor in the skyrmion case (right scale). These results prove independent of the damping parameter provided the time step in the integration of the LLG equation be suitably chosen.}
\label{Fig_WSk_Diff_14} 
\end{figure}

One certainly expects from the simulation model a fair prediction of the average magnetization $\langle M_z \rangle$ or $\langle S_z \rangle$  \textit{vs}  temperature $T$, at least for temperatures substantially lower than the Curie temperature $T_C$.  Fig.\ref{Fig_WSk_Diff_14} shows the variation of $\langle M_z \rangle / M_z(T=0)$ or $\langle S_z \rangle / S_z(T=0)$ with temperature for the two model magnetic layers of this work. Although simulation results do not compare unfavorably with published experimental data \cite{Shimamura:2012, Koyama:2015, Obinata:2015}, where, typically, the Curie temperature amounts to $\approx 150 K$ for 1 ML, and proves larger than $300 K$ for thicknesses above 2 ML, a more detailed analysis, potentially including disorder, ought to be performed. 
\begin{equation}
\label{eq:WSkDiff:20}
\begin{split}
\langle G_z \rangle &= \frac{\mu_0  \mu_{At} \langle s_z \rangle}{\gamma_0}  \langle \sum_k
\left[  \vec{s}(k) \cdot  \left[ \partial_x  \vec{s}(k) \times  \partial_y \vec{s}(k) \right] \right]  \rangle
\\
& = \frac{\mu_0  \mu_{At} \langle s_z \rangle}{\gamma_0 S_{At}} \langle G_z^{\mathrm{VF}} \rangle 
\\
\langle D_{xx} \rangle  &=  \frac{\mu_0  \mu_{At} \langle s_z \rangle}{\gamma_0} \langle { \sum_k \left[ \partial_x \vec{s}(k) \right]^2 } \rangle
\\
&=  \frac{\mu_0  \mu_{At} \langle s_z \rangle}{\gamma_0 S_{At}} \langle D_{xx}^{\mathrm{VF}} \rangle
\end{split}
\end{equation}

Let us now, without further justification,  substitute in the expression of the skyrmion diffusion coefficient time averaged values of $G$ and $D$, owing to relations (\ref{eq:WSkDiff:20}). Keeping in mind the geometrical meaning of $G_z^{\mathrm{VF}}$, the dimensionless vector function in $G$, $\langle G_z \rangle$ is anticipated to be a sole function of $\langle s_z \rangle$. Inversely, $D_{xx}^{\mathrm{VF}}$, the (dimensionless) vector function in $\langle D_{xx} \rangle$, a definite positive quantity, steadily increases with thermal disorder. It is even found to be proportional to temperature (not shown). Its time averaged value for the sole skyrmion may only be obtained by subtraction of values computed in the presence and absence of the skyrmion.

For the skyrmion in our model Co monolayer, $\langle D_{xx}^{\mathrm{VF}} \rangle$ is found to increase moderately with temperature (see Fig.\ref{Fig_WSk_Diff_14}), a result also anticipated from an increase with temperature of the skyrmion radius. Besides, both $\langle G_z \rangle$ and $\langle D_{xx} \rangle$ are expected to decrease with temperature due to their proportionality to  $\langle s_z \rangle$.  $\langle D_{xx} \rangle$ is thus subject to two competing effects of temperature $T$. Present evidence, however, points at a dominating influence of  $\langle s_z(T) \rangle$.

\section{\label{Summary}Summary and Outlook\protect\\ }

Summarizing, it has been shown that the Brownian motion of chiral walls and skyrmions in DMI materials obeys diffusion equations with markedly different damping parameter ($\alpha$) dependence. Although not a new result, skyrmions Brownian motion suppression with decreasing $\alpha$ ($\alpha < G/D$) is substantiated by a wide exploration of the damping parameter space. The observation of this astonishing topological property might, however, be hampered by the restriction to ultra-small skyrmion sizes or ultra-low $\alpha$ values. The discrepancy (up to 20\%) between simulation results and theoretical expectations could be reduced by the introduction of time averaged values for the gyrotropic and dissipation contributions to the analytical diffusion coefficients in the "low" noise limit, at the expense of a tiny upwards curvature in the $\mathcal{D}(T)$ curves. A strong theoretical justification for doing so remains, however, lacking at this stage.

In this work, the sample has been assumed to be perfect, i.e. devoid of spatial variations of the magnetic properties, even though the lifting of such a restriction is anticipated to prove mandatory for a proper description of experiments. Diffusion in the presence of disorder has been theoretically studied for a number of disorder and random walk types  \cite{Bouchaud:1990, Metzler:2000}. Generally, disorder changes the linear growth with time of the position variance into a power law, a behavior called superdiffusion if the exponent is larger than 1 and subdiffusion if smaller. For instance, if the skyrmion motion in a disordered system may be mapped onto a 2D random walk with an onsite residence time $\tau$, probability $\propto \tau^{-(1+\mu)}$ ($\mu < 1$), then the diffusion exponent will be $\mu$, meaning subdiffusion. Besides, choosing a physically realistic disorder model for a Co monolayer might well prove equally arduous \cite{Meier:2006}. Altogether, skyrmion diffusion in the presence of disorder has been left out for future work.      

\begin{acknowledgments}
Support by the Agence Nationale de la Recherche (France) under Contracts
No. ANR-14-CE26-0012 (Ultrasky), No. ANR-17-CE24-0025 (TopSky) is gratefully acknowledged.
\end{acknowledgments}
%
%
%
%
%
%\bibliography{Wall&Skyrmion_Diff_V2_ReSubm}
% Produces the bibliography via BibTeX.

%Merlin.mbs v4.21 2009-07-09.
\providecommand{\noopsort}[1]{}\providecommand{\singleletter}[1]{#1}%

\end{document}